\documentclass[smallcondensed]{svjour3}

\usepackage{amsmath}
\usepackage{graphicx}
\usepackage{subfig}
\usepackage{url}
\usepackage{hyperref}
\hypersetup{
 colorlinks=true,
 linkcolor=blue,
 filecolor=blue,
 citecolor = black,      
 urlcolor=cyan,
}

\usepackage{color}
\usepackage{etoolbox}
\usepackage{amsmath,amssymb,amsfonts}

\usepackage{mathtools}
\usepackage{tabularx}
\usepackage{textcomp}
\usepackage{multirow}
\usepackage{multicol}
\usepackage{color,soul}
\setstcolor{red} 
\usepackage{natbib}
\usepackage{makecell}

\usepackage{pdflscape} 
\usepackage{pdfpages}
\usepackage{rotating}

\allowdisplaybreaks

\usepackage{caption}

\usepackage{booktabs}
\usepackage{tikz}
\usetikzlibrary{arrows,positioning}
\usepackage{lipsum, adjustbox}
\usepackage{algorithm2e}

\usepackage{setspace}

\usepackage{xcolor}

\newcommand{\mcps}{\multicolumn{1}{p{2.09cm}}}

\newcommand{\cnt}{\centering}
\newcommand{\rrt}{\raggedright}

\begin{document}\sloppy

\title{Improved $\alpha$-GAN architecture for generating 3D connected volumes with an application to radiosurgery treatment planning}
\titlerunning{Improved $\alpha$-GAN architecture for generating 3D connected volumes}

\author{
Sanaz Mohammadjafari\and
Mucahit Cevik \and
Ayse Basar
}

\institute{Sanaz Mohammadjafari \at
Data Science Lab, Toronto Metropolitan University, Toronto, Canada\\
\email{sanaz.mohammadjafari@ryerson.ca}\\[-1.0em]
\and
Mucahit Cevik \at 
Data Science Lab, Toronto Metropolitan University, Toronto, Canada\\[-1.0em]
\and
Ayse Basar\at 
Data Science Lab, Toronto Metropolitan University, Toronto, Canada\\[-1.0em] 
}

\date{Received: date / Accepted: date}

\maketitle

\begin{abstract}
Generative Adversarial Networks (GANs) have gained significant attention in several computer vision tasks for generating high-quality synthetic data. Various medical applications including diagnostic imaging and radiation therapy can benefit greatly from synthetic data generation due to data scarcity in the domain. However, medical image data is typically kept in 3D space, and generative models suffer from the curse of dimensionality issues in generating such synthetic data.
In this paper, we investigate the potential of GANs for generating connected 3D volumes. We propose an improved version of 3D $\alpha$-GAN by incorporating various architectural enhancements. On a synthetic dataset of connected 3D spheres and ellipsoids, our model can generate fully connected 3D shapes with similar geometrical characteristics to that of training data. We also show that our 3D GAN model can successfully generate high-quality 3D tumor volumes and associated treatment specifications (e.g., isocenter locations).
Similar moment invariants to the training data as well as fully connected 3D shapes confirm that improved 3D $\alpha$-GAN implicitly learns the training data distribution, and generates realistic-looking samples. The capability of improved 3D $\alpha$-GAN makes it a valuable source for generating synthetic medical image data that can help future research in this domain.

\end{abstract}

\section{Introduction}

Over the past decade, Generative Adversarial Networks (GANs) have shown promising results in modeling a distribution from the available samples  \citep{goodfellow2014generative}. GANs have been frequently employed for image-to-image translation \citep{lee2018diverse}, image super-resolution \citep{wang2018high} and specifically for synthesizing real-world images from random noise inputs \citep{zhong2019high,brock2018large,mohammadjafari2021designing}.
GANs consist of two competing networks, namely, a generator that generates similar samples to the true distribution, and a discriminator that differentiates between the training and generated samples. GANs' goal is to generate novel and diverse samples which are representative of the real data distribution. 

The majority of the recent works on GANs are typically confined to 2D datasets. Increasing datasets' dimensionality significantly impacts the computational complexity of the generative models. However, the development of powerful computing architectures and decreasing costs of computational power have made it possible to work with 3D images. Although moving from 2D to 3D images increases the computational burden, it provides a more detailed representation of the instances in the dataset. The impact of this transition is highlighted in the medical domain, where 3D images offer a more detailed view of human organs that are crucial to detect abnormalities \citep{suzuki2017overview}. 
We note that many recent studies on 3D GANs have focused on generating 3D objects from available 2D views. For instance, \citet {chan2021pi} and \citet{or2022stylesdf} employed an encoder structure to extract features from the available 2D data and used the learned latent space for 3D image generation. On the other hand, training instability and mode collapse issues in GANs' training procedure contribute to the lack of research on generating 3D images directly from random noise vectors.

GANs' potential capabilities in generating high-quality 3D synthetic images have inspired various real-life applications, particularly in the medical domain, where the available data is limited due to privacy concerns, and publicly available data is rarely found. 
For instance,  \citet{han2019synthesizing} used GANs to generate 3D lung nodules' 
Computed Tomography (CT) samples for data augmentation and enhanced the performance of object detection. Another important application of 3D GANs is generating treatment plans as data samples for knowledge-based treatment
planning \citep{mahmood2018automated}. Automation in radiation therapy treatment planning is highly desirable as the radiological images require image analysis and diagnosis by a trained human radiologist. 
That is, the high cost of training radiologists and the time-consuming procedure of manual diagnosis make machine learning models good candidates for assisting clinicians to reduce the workload. One specific use case for synthetic 3D image generation is radiosurgery treatment planning, where the objective is to find a set of isocenters (i.e., focus points for radiation beams) and associated radiation amounts so that tumor volumes are eradicated with minimum dosage to surrounding healthy tissues. \citet{berdyshev2020knowledge} proposed ResNet models to learn from existing brain tumor data and previous treatment plans, and the authors noted the need for larger datasets for a high-performance automated ML-based treatment planning. 
Additionally, data privacy concerns prevent making such datasets publicly available, and synthetic data generation can offer a remedy for this issue.

\paragraph{\textbf{Research goal}}
We propose an improved version of 3D $\alpha$-GAN architecture to generate 3D connected volumes from noise vectors. Although recent works have explored the generation of 3D objects, they fall short of addressing the connectivity requirements for the generated pixels. Our proposed model benefits from inception blocks inside the discriminator's network among other modifications. The small and variant sizes of kernels in the inception block help the discriminator with identifying connectivity in data. Moreover, we evaluate our model on four synthetically generated datasets and extend the evaluation to synthetic data generation for radiosurgery treatment planning.

\paragraph{\textbf{Contributions}}
The main contribution of our paper is the design of an improved 3D $\alpha$-GAN structure for the generation of 3D connected volumes. 
Specific contributions of our study can be summarized as follows:

\begin{itemize}\setlength\itemsep{0.3em}
\item We modify the 3D $\alpha$-GAN \citep{kwon2019generation} architecture by introducing a new discriminator that incorporates the inception blocks. The diverse set of kernels in the inception block helps to extract fine details such as connectivity. As such, our proposed architecture constitutes a novel adaptation of 3D $\alpha$-GAN. 

\item We focus on the connectivity considerations between an object's pixels and design a set of evaluation metrics to assess the connected components in 3D space. To the best of our knowledge, this is the first study to investigate connectivity considerations in 3D image generation.

\item We propose an extensive set of evaluation metrics to estimate the shape and proper distribution of 3D volumes and the artifacts within (e.g., isocenters in radiosurgery data). We also conduct a thorough numerical study to evaluate different GAN variants' performance.

\item Our results offer a new insight into 3D object generation using voxelized representations of tumor volumes. Accordingly, our analysis provides evidence for the potential of GANs in synthetic data generation for radiosurgery treatment planning. 
\end{itemize}

\paragraph{\textbf{Organization of the paper}}
The rest of this paper is organized as follows. Section~\ref{sec:Background} summarizes the recent developments in 3D object generation using GANs. In Section~\ref{sec:Methodology}, generic Variational Auto-encoder (VAE) and GAN structures are presented, and details of the proposed architecture are explained. Section~\ref{sec:results} provides a comprehensive list of our proposed evaluation metrics as well as a discussion on the model performance for four different datasets. Finally, the paper is concluded in Section~\ref{sec:conclusion} along with a summary of our findings and a discussion of potential future research directions.

\section{Background}\label{sec:Background}
In this section, we review recent studies on 3D object synthesis and discuss various GANs' architectures focused on generating novel, diverse 3D shapes. Moreover, we explore the applications of 3D object generation in radiosurgery treatment planning. A summary of the most closely related studies on 3D GANs is reported in Table~\ref{tbl:lit_rev_ch3}, which contains information on the proposed methodologies and employed datasets. 

\setlength{\tabcolsep}{7pt}
\renewcommand{\arraystretch}{1.5}
\begin{table}[!ht]
\caption{Summary of the relevant papers on 3D object generation}
\label{tbl:lit_rev_ch3}
\resizebox{0.99\textwidth}{!}{
\begin{tabular}{p{0.25\textwidth} p{0.18\textwidth} p{0.37\textwidth} p{0.37\textwidth}}
\toprule
{\bf Study} & {\bf Model} & {\bf Methodology} & {\bf Datasets}\\
\midrule
\citet{wu2016learning} & 3D-GAN,\hspace{0.2cm} 3D-VAE-GAN &  Convolutional GAN combined with VAE & ModelNet, Ikea dataset\\
\midrule
\citet{choy20163d} & 3D-R2N2 & Learned mapping from 2D images to generate 3D objects &  ModelNet, PASCAL VOC$^1$, Online Products$^2$ \\
\midrule
\citet{kwon2019generation} & 3D $\alpha$-GAN & Adaptation of $\alpha$-GAN plus WGAN-GP  & ADNI$^3$ , BRATS 2018$^4$ \\
\midrule
\citet{babier2020knowledge} & 3D GAN & 3D pix-to-pix structure &  Clinical radiation therapy plans for 217 patients with oropharyngeal cancer\\

\midrule
\citet{hong20213d} & 3D-StyleGAN & Mapped latent noise with style vector  & Brain MR T1 images\\
\midrule
\citet{jangid20223d} & M-GAN & Adaptation of StyleGAN  & ModelNet, 3D grain volumes\\

\midrule
Our study &  Improved 3D $\alpha$-GAN & Adaptation of 3D $\alpha$-GAN with inception-layered discriminator & Synthetic connected 3D sphere/ellipsoid volumes with/without packed spheres, Synthetic connected 3D tumor volumes with/without packed spheres\\
\bottomrule
\end{tabular}
}
\\
{\scriptsize $^1$: from \citep{everingham2011pascal}}\\
{\scriptsize $^2$: from \citep{oh2016deep}}\\
{\scriptsize $^3$: from \url{https://adni.loni.usc.edu/}}\\
{\scriptsize $^4$: from 
\citep{bakas2017advancing}}
\end{table}

3D object generation and detection are one of the most important research areas in computer vision \citep{xiang2016objectnet3d,zhi2018toward}.
Three-dimensional images are typically represented in four formats, including point cloud, voxel grid, triangle mesh, and multi-view representation. Voxels are 3D versions of image pixels and can be interpreted as quantized, fixed-sized point clouds. On the other hand, point clouds have an infinite number of points in space with float pixel coordinates. Voxel grids suffer from large memory consumption and low-resolution representation. However, in certain problems (e.g., radiotherapy) the limited size of 3D shapes and lack of fine details make the 3D voxel grid a suitable representation for coarse objects. 

There has been a dramatic increase in 3D object modeling and synthesis in recent years, with several studies focusing on incorporating the existing objects' parts in CAD model libraries. 
These approaches generate realistic-looking samples with low novelty and diversity \citep{van2011survey}. Developments in deep learning and the introduction of large 3D CAD datasets such as ModelNet \citep{wu20153d} have attracted significant attention toward learning deep representations from voxelized objects. 
It is important to note that using deep learning models on 3D objects is more difficult than on 2D objects because of the curse of dimensionality. 
\citet{wu20153d} were the first to represent 3D shapes as probabilistic distribution of binary voxels, where ones represent data, and zeros represent empty space. They introduced a Convolutional Deep Belief Network (CDBN) called ShapeNet, which is the first deep learning model to be trained on 3D voxel objects. Drawing on this study, several methods were proposed to incorporate deep learning models in 3D image-related tasks such as 3D object recognition \citep{maturana2015voxnet}, 3D object representation and retrieval using 2D views \citep{li2015joint,jimenez2016unsupervised, smith2018multi} and 3D image reconstruction from noisy data \citep{sharma2016vconv}.  


The introduction of adversarial loss to the GAN models as well as the discriminator's ability in object recognition inspired the incorporation of GANs for synthetic 3D object generation. 
\citet{wu2016learning} were the first to propose 3D-GAN and 3D-VAE-GAN for synthesizing diverse and high-quality 3D objects. 
Their trained discriminator outperformed the accuracy of state-of-the-art models for image classification over ModelNet dataset by a margin of 15\%. 
However, the authors did not offer any evaluation metrics for the quality of generated samples. 
Following this work, \citet{choy20163d} proposed a 3D Recurrent Reconstruction Neural Network (3D-R2N2) to learn a mapping from multi-view 2D images to their 3D shapes.
This work led to a large number of studies that focused on reconstructing a 3D object from its 2D slices, leveraging the available information in 2D representation \citep{zhang2018learning,wu2017marrnet}. 
However, using 2D representation limits the generation of novel and diverse samples as it confines the latent space of images. 
\citet{hong20213d} proposed an extended version of 2D-StyleGAN \citep{karras2020analyzing}, called 3D-StyleGAN, to generate 3D MRI scans of brain.
Their architecture has a progressive structure, where each part receives a version of the mapped latent noise vector and a style vector and passes them through 3D convolutions. 
The model generates high-quality 3D brain scans that can be manipulated using the style vectors. 
Similarly, \cite{jangid20223d} proposed M-GAN by adapting the StyleGAN \citep{karras2019style} to generate 3D grain shapes in Polycrystals. 
They noted that, although their model generates reasonable small 3D grain volumes, grain connectivity relationships are not guaranteed, and should be investigated. 
Following the same procedure, \citet{kwon2019generation} suggested an adaptation of $\alpha$-GAN \citep{rosca2017variational} for 3D space by adding the Wasserstein GAN with Gradient Penalty (WGAN-GP) \citep{gulrajani2017improved} loss functions. 
The proposed model uses a combination of GANs and VAEs to address the mode collapse and image blurriness.  

3D object generation has always been an important problem in the medical domain as the majority of available datasets are 3D, and the images typically have significantly more details. 
The computational complexity and memory issues commonly led to using an alternative approach by extracting 2D slices of the 3D images to address this issue, causing a loss of information and connectivity between the slices \citep{volokitin2020modelling,mahmood2018automated}. 
The GANs' capabilities in synthesizing 3D images have increased their potential in medical applications, such as generating more diverse and unseen 3D brain MRI data to address clinically difficult tasks \citep{hong20213d, chong2021synthesis, kwon2019generation}. Another significant application of GANs in the medical domain is generating treatment planning schemes to help with knowledge-based planning and treatment planning automation \citep{berdyshev2020knowledge, cevik2018modeling}. 


\citet{babier2020knowledge} proposed a 3D GAN image-to-image translation structure adapted from Pix2pix \citep{isola2017image} to solve a dose distribution problem. Their model receives CT images and predicts the full 3D-dose distribution. Their approach increases the satisfied clinical criteria by 11\% compared to other clinical approaches. Additionally, their method outperforms a similar 2D GAN structure applied on 2D slices of the image \citep{mahmood2018automated}, confirming that the correlations between 2D slices offer useful information. \citet{berdyshev2020knowledge} specifically focused on predicting the locations of isocenters inside the tumor volumes, which is an important step in radiosurgery treatment planning. They proposed a dimensionality reduction technique for the 3D tumor shapes to avoid the curse of dimensionality and employed a residual neural network for the isocenter prediction task. Although they showed comparable results to the ground truth in their limited experimental results, their approach lead to the loss of voxel connectivity by moving to 2D space. 


%

These studies inspire the investigation of the capabilities of 3D GANs in radiosurgery treatment planning.
Our proposed architecture is an adapted version of 3D $\alpha$-GAN to generate connected 3D volumes. 
Our model differs from 3D $\alpha$-GAN \citep{kwon2019generation} as we adopt an inception structure for our discriminator network, and incorporate a specifically designed loss function to ensure connectivity.





\section{Methodology}\label{sec:Methodology}
In this section, we explain the details of our proposed architecture, datasets, and experimental setup. 
We first review the basics of GAN and VAE and then explain the improved 3D $\alpha$-GAN architecture in detail. 
Next, we discuss the specifications of four synthetic datasets used in our analysis and provide the details of our experimental setup. Note that, for the sake of clarity, we provide a notation table in Section~\ref{ap:notations} of Appendix. 

\subsection{Review of GAN models}
\begin{figure}[!ht]
\centering
\includegraphics[width=0.85\textwidth]{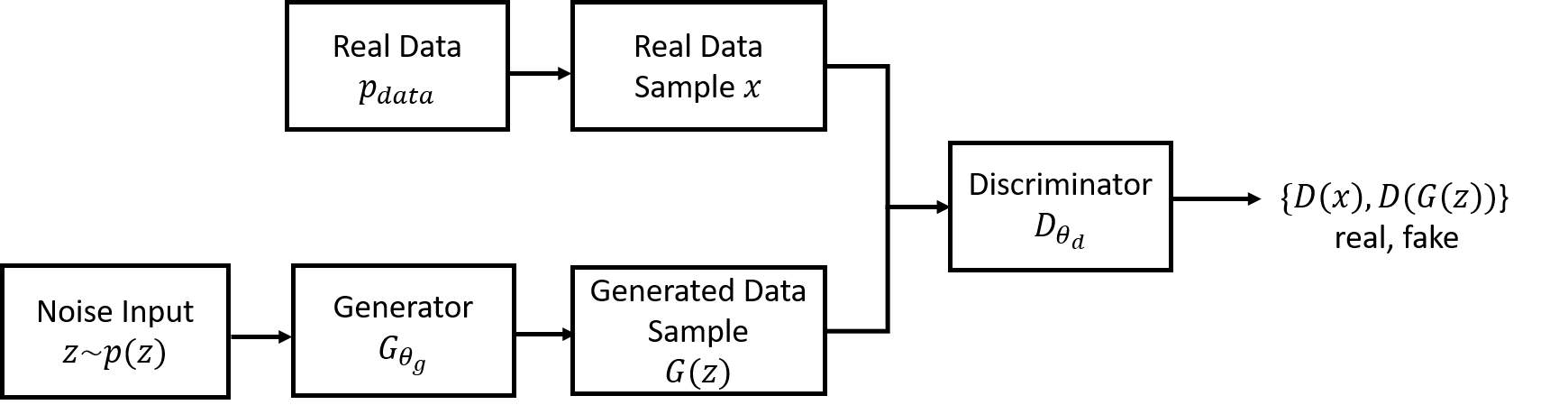}
\caption{Vanilla GAN structure that consists of two competing networks, namely, discriminator and generator}
\label{fig:GAN_structure}
\end{figure}

A generic GAN architecture contains two networks known as the generator and the discriminator, both competing against each other (see Figure~\ref{fig:GAN_structure}). 
During the training process, the generator learns the distribution $p_g$ from an input noise $z$ mapped through the function $G(z; \theta_g)$ to the samples. 
Here, $G(z; \theta_g)$ is a differentiable function, usually defined as a neural network with a set of parameters $\theta_g$. 
The discriminator $D(x,\theta_d)$, which is also a differentiable function with parameters $\theta_d$, maps input samples to a probability value indicating whether the sample is real or generated. 
Both networks are trained simultaneously, where the discriminator and generator have maximization and minimization objectives, respectively.
Therefore, GAN is modeled as a min-max game with a value function $V(D,G)$, that is,
\begin{equation}
     \label{eq:GANS-formula}
    \min_G{\max_D{V(D,G)}} = E_{x\sim p_{data}(x)}[\log D(x)] + E_{z\sim p_{z}(z)}[\log(1-D(G(z)))]
\end{equation}


The discriminator aims to maximize the value function by generating a probability value of one for the samples from the real data, $p_{data}$, and zero for the samples from the generated distribution, $p_g$. 
On the other hand, generator minimizes the objective value by generating real-looking samples that the discriminator fails to detect as fake.



\subsection{Variational auto-encoders}
\begin{figure}[!ht]
\centering
\includegraphics[width=1\textwidth]{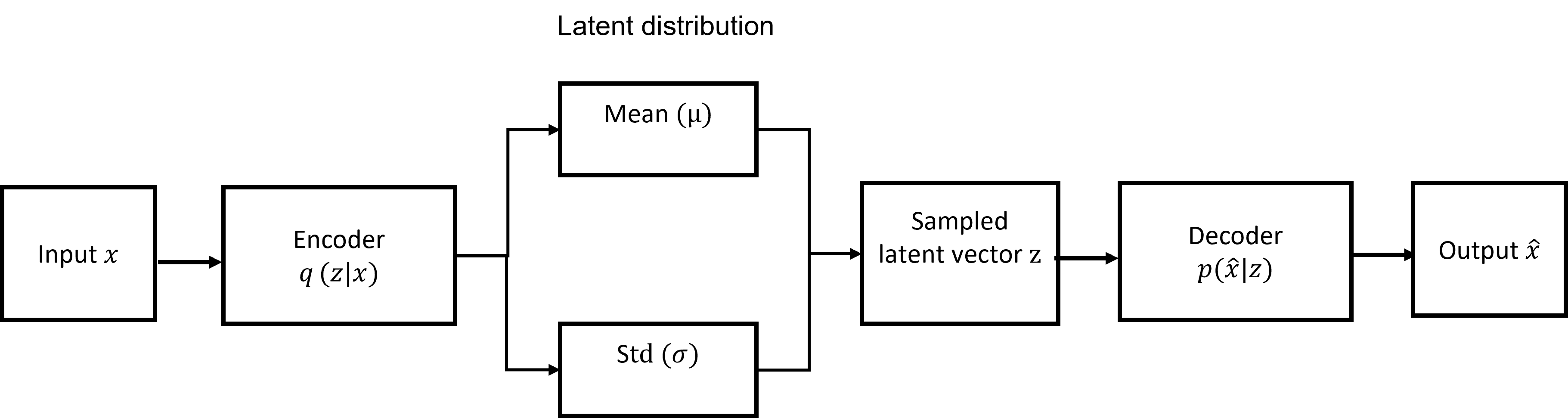}
\caption{VAE structure consists of two networks, called, encoder and decoder}
\label{fig:VAE}
\end{figure}

VAE is a generative model with a similar structure as a regular auto-encoder that uses variational inference for training and creates a distribution for each attribute of the latent space instead of a single value. 
VAE consists of two networks, namely, encoder and decoder, as illustrated in Figure~\ref{fig:VAE}. 
The encoder maps the input $x$ to a continuous smooth latent space $z$, and decoder samples from the latent space distribution to generate diverse samples at the output. 
The trained decoder can be used as a generator to create diverse samples from the latent space distribution. 
Since the random sampling process cannot generate gradients during the backpropagation and model's training process, a re-parameterization trick was proposed to address this issue \citep{DBLP:journals/corr/KingmaW13}. 
We assume that the prior distribution $p(z)$ follows a Gaussian distribution, thus the latent space presents two vectors that contain the mean and standard deviation of the encoded distribution. 

The VAE's loss function is shown in Equation~\eqref{eq:VAE_loss}, and it consists of a reconstruction error between the input and generated image and a regularization term for latent space samples $j$ that follows a Gaussian distribution.  
\begin{equation}
     \label{eq:VAE_loss}
    L= \mathcal{L}(x,\hat{x})+\sum_j [\mathbb{KL}(q_j(z|x)||p(z))] 
\end{equation}
Since VAEs use the encoded space of real samples, unlike GANs, they do not suffer from mode collapse. 

\subsection{3D $\alpha$-GAN architecture}\label{sec:3d_alpha_GAN}
\begin{figure}[!ht]
\centering
\includegraphics[width=1\textwidth]{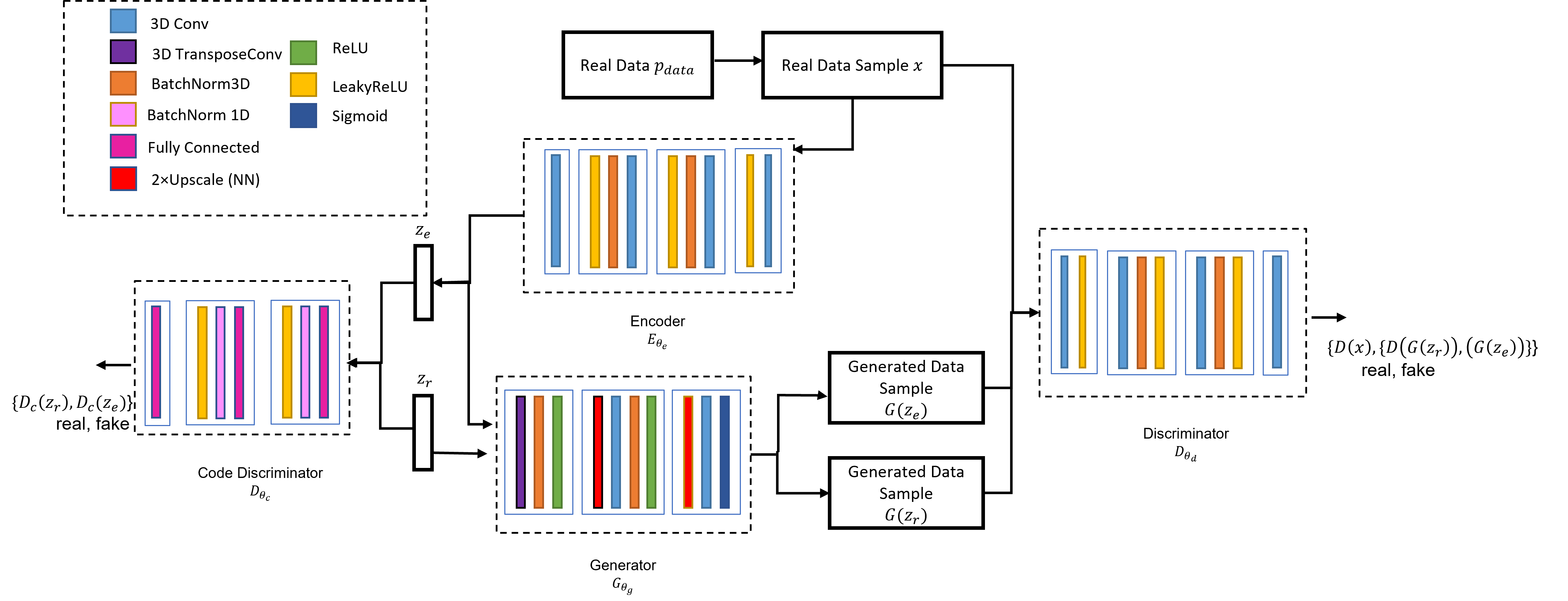}
\caption{$\alpha$-GAN structure consists of four competing networks, namely, discriminator, code discriminator, encoder and generator}
\label{fig:Alpha_WGAN_with_networks}
\end{figure}

We leverage the 3D $\alpha$-GAN structure proposed by \citet{kwon2019generation}, which is an  adaptation of $\alpha$-GAN architecture \citep{rosca2017variational}. 
$\alpha$-GAN is a combination of VAE and GAN to solve the mode collapse issue in GANs and overcome the blurry images generated by VAEs. 
It consists of four networks, namely, generator $G(z; \theta_g)$, discriminator $D(x; \theta_d)$, encoder $E(x; \theta_e)$ and code discriminator $D_c(x; \theta_c)$. 
Code discriminator network replaces the variational inference in VAE.
The encoder receives the real data samples and encodes them to a latent vector $z_e$. 
The encoded vector $z_e$, as well as the noise vector $z_r$, are both passed to the generator and code discriminator. 
The generator in this approach generates two sets of fake data from encoded noise vector $z_e$ and predefined noise vector $z_r$ and passes them to the discriminator. 
The discriminator receives two generated samples and one real sample and differentiates between fake and real samples. 
On the other hand, the code discriminator distinguishes between the real noise vector $z_r$ and the encoded one $z_e$. 
The code discriminator encourages the encoder to fully encode the real distribution to $z_e$. When the code discriminator fails to differentiate between fake and real noise, it shows that the generator has covered the entire decoded space, and the network has successfully prevented the mode collapse. Moreover, when the discriminator fails to recognize the fake from real samples, the network is fully trained, and the generated samples are representative of the training distribution. Figure~\ref{fig:Alpha_WGAN_with_networks} illustrates the model architecture and networks' structures. 

\citet{kwon2019generation} showed that replacing the basic GAN's loss function using Wasserstein loss with gradient penalty \citep{gulrajani2017improved} stabilizes the model's training process. Equation~\eqref{eq:D_Loss_alphaGAN_WGAN_formula}presents the discriminator's loss function of a general GAN plus the gradient penalty factor formulated as $L_{\texttt{GPD}} = E_{\hat{x}}[(||\nabla_{\hat{x}}D(\hat{x})||_2-1)^2]$. 
\begin{equation}
     \label{eq:D_Loss_alphaGAN_WGAN_formula}
    L_D= E_{z_e}[D(G(z_e))] + E_{z_r}[D(G(z_r))] - 2E_{x_{real}}[D(x_{real})] + \lambda_1 L_{\texttt{GPD}}
\end{equation}
Loss functions of generator and encoder are equivalent, and represented in Equation~\eqref{eq:G_Loss_alphaGAN_WGAN_formula}, where the reconstruction loss is added using the $L_1$ distance between the real and generated image, that is,
\begin{equation}
     \label{eq:G_Loss_alphaGAN_WGAN_formula}
    L_G= -E_{z_e}[D(G(z_e))] - E_{z_r}[D(G(z_r))] + \lambda_2||x_{real}-G(z_e)||_{L_1}
\end{equation}
The gradient penalty of code discriminator is defined by $L_{\texttt{GPC}}$ and formulated as $E_{z}[(||\nabla_{z}D_c(z)||_2-1)^2]$. We can obtain the code discriminator loss and its gradient penalty term as
\begin{equation}
     \label{eq:C_Loss_alphaGAN_WGAN_formula}
     L_C= E_{z_e}[D_c(z_e)] - E_{z_r}[D_c(z_r)] + \lambda_1 L_{\texttt{GPC}}
\end{equation}
Note that $\lambda_1$ and $\lambda_2$ are predefined hyperparameters, which we set to 10 in our implementations.


\subsection{3D $\alpha$-GAN with connection loss}
\begin{figure}[!ht]
\centering
\subfloat[A sample of one connected object {(shape on the left has a \texttt{CC} of $[2]$, whereas the shape on the right has a \texttt{CC} of $[1]$)} \label{fig:Connection_loss_1}]{\includegraphics[width=0.45\textwidth]{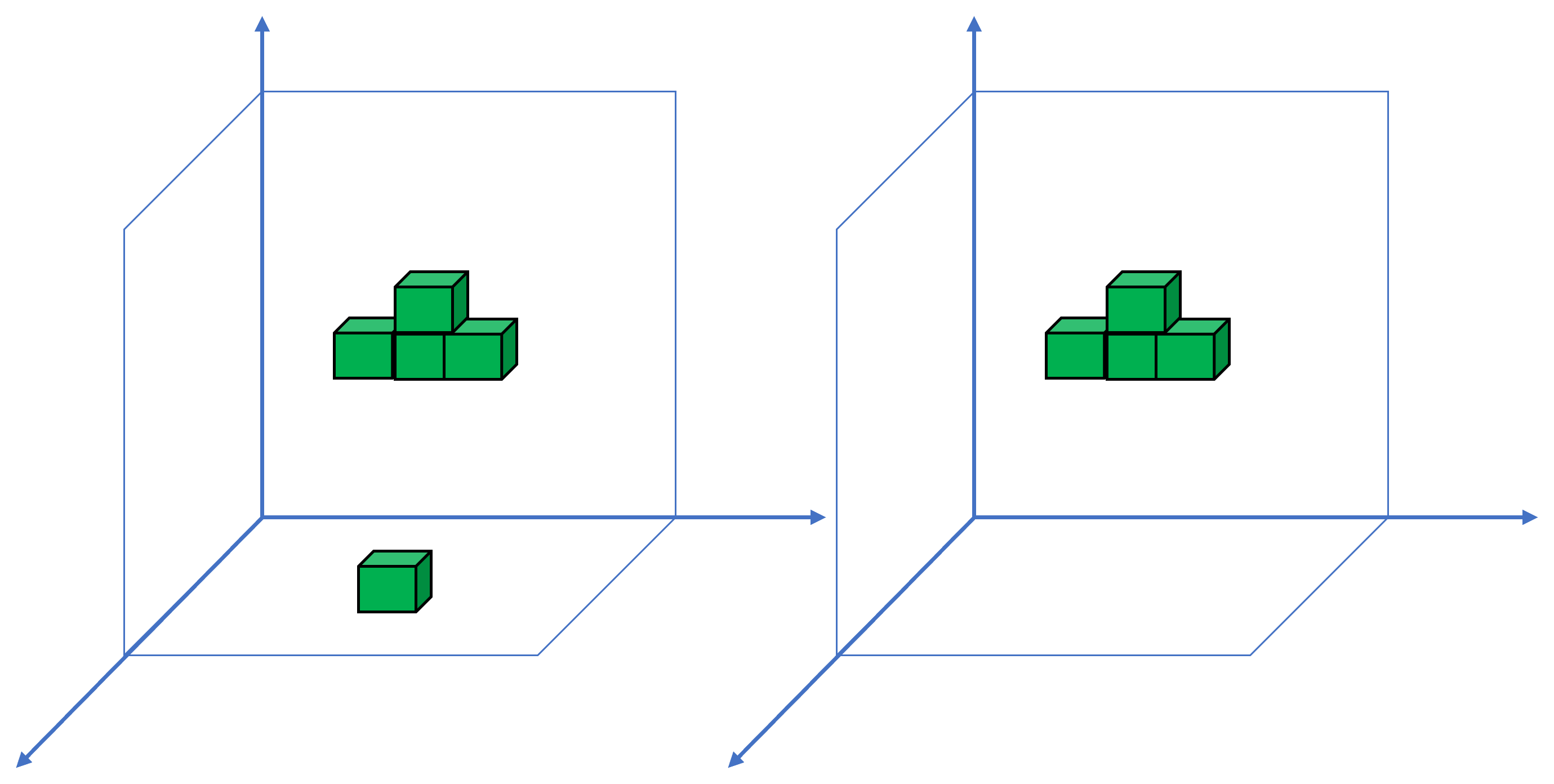}}
\hfill
\subfloat[A sample of two sets of connected objects {(shape on the left has a \texttt{CC} of $[1,1]$, whereas the shape on the right has a \texttt{CC} of $[1,2]$)} \label{fig:Connection_loss_2}]{\includegraphics[width=0.45\textwidth]{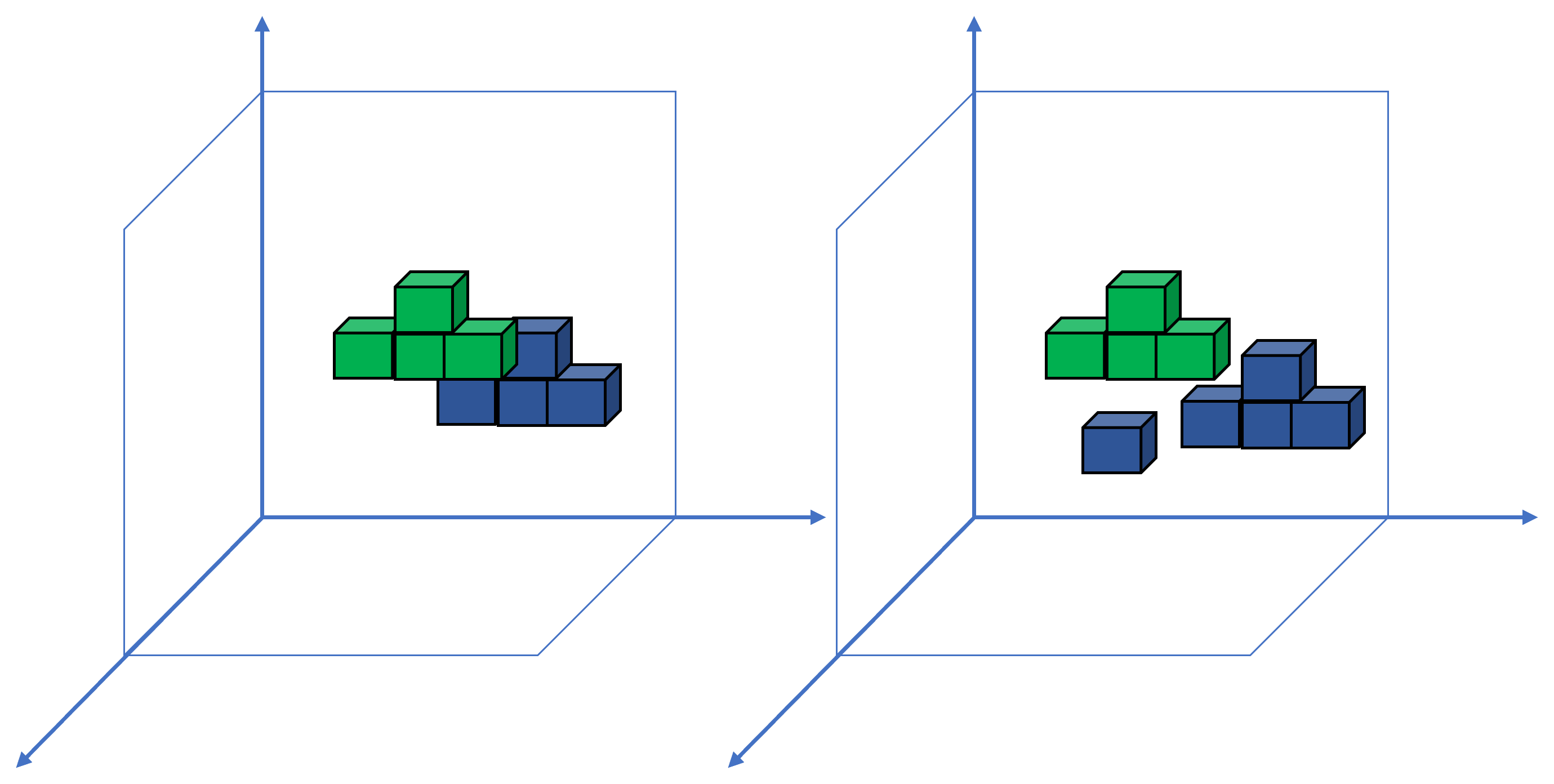}}
\caption{Connected components of two different samples}
\label{fig:Connection_loss}
\end{figure}

The existing $\alpha$-GAN architecture does not provide any insights regarding the connection of elements in 3D objects. 
We make a small alteration in the generator's loss function that can significantly improve the model's performance in generating connected components. 
Let the number of connected components in an object be $\texttt{CC}$, which should ideally be equal to one. 
We define the connection loss as 
\begin{equation}
\begin{small}
\label{eq:L_connect_loss}
     L_{connect}=
     \begin{cases}
     \displaystyle{\frac{1}{N}}\displaystyle{\sum_{i=1}^{N}}{(|\texttt{CC}-1|-\lambda_3[\texttt{CC}=1])} & \# \textit{connected objects} = 1\\
  \displaystyle{\frac{1}{N}}\displaystyle{\sum_{i=1}^{N}}{(\sqrt{\displaystyle{\frac{1}{K}}\displaystyle{\sum_{i=1}^{K}}(\texttt{CC}[i] - 1)^2} - \lambda_3 \displaystyle{\sum_{i=1}^{K}}{[\texttt{CC}[i] = 1]})} & \# \textit{connected objects} \geq 2
    \end{cases}
\end{small}
\end{equation}
Equation~\eqref{eq:L_connect_loss} shows an additional penalty term, which is added to the generator's loss function to encourage the generation of more connected elements.
For samples with one connected 3D object, loss for $N$ samples is calculated over the number of connected elements. 
On the other hand, when the shape includes $K$ connected elements, the root mean square error between one and the number of connected components is calculated and added to a hyperparameter $\lambda_3$ multiplied by the number of connected components that are equal to one.
Here, we set the initial value for $\lambda_3$ as one, but the hyperparameter tuning process could provide a better alternative value. 
The first element of the loss penalizes the generator for disconnected components, whereas the second part of the loss provides a large reward for outputting fully connected components. 
Note that the second term is included to not overwhelm the generator at the initial training epochs. 
Samples of connected components are illustrated in Figure~\ref{fig:Connection_loss} to provide a more detailed explanation on connection loss computation. For Figure~\ref{fig:Connection_loss_1} with one connected object and two samples, the connection loss is computed as $\frac{1}{2}[(|2-1|-1\times0) + (|1-1|-1\times1)]$. On the other hand, for Figure~\ref{fig:Connection_loss_2} with two connected objects and two samples, the connection loss is estimated as $\frac{1}{2}[(\sqrt{\frac{1}{2}[(1-1)^2 + (1-1)^2]}-1\times2) + (\sqrt{\frac{1}{2}[(1-1)^2 + (2-1)^2]}-1\times1)]$.
\subsection{Improved 3D $\alpha$-GAN architecture}
\begin{figure}[!ht]
\centering
\subfloat[$\alpha$-GAN structure with the new proposed discriminator \label{fig:improved_Alpha_WGAN_with_networks_dis}]{\includegraphics[width=1\textwidth]{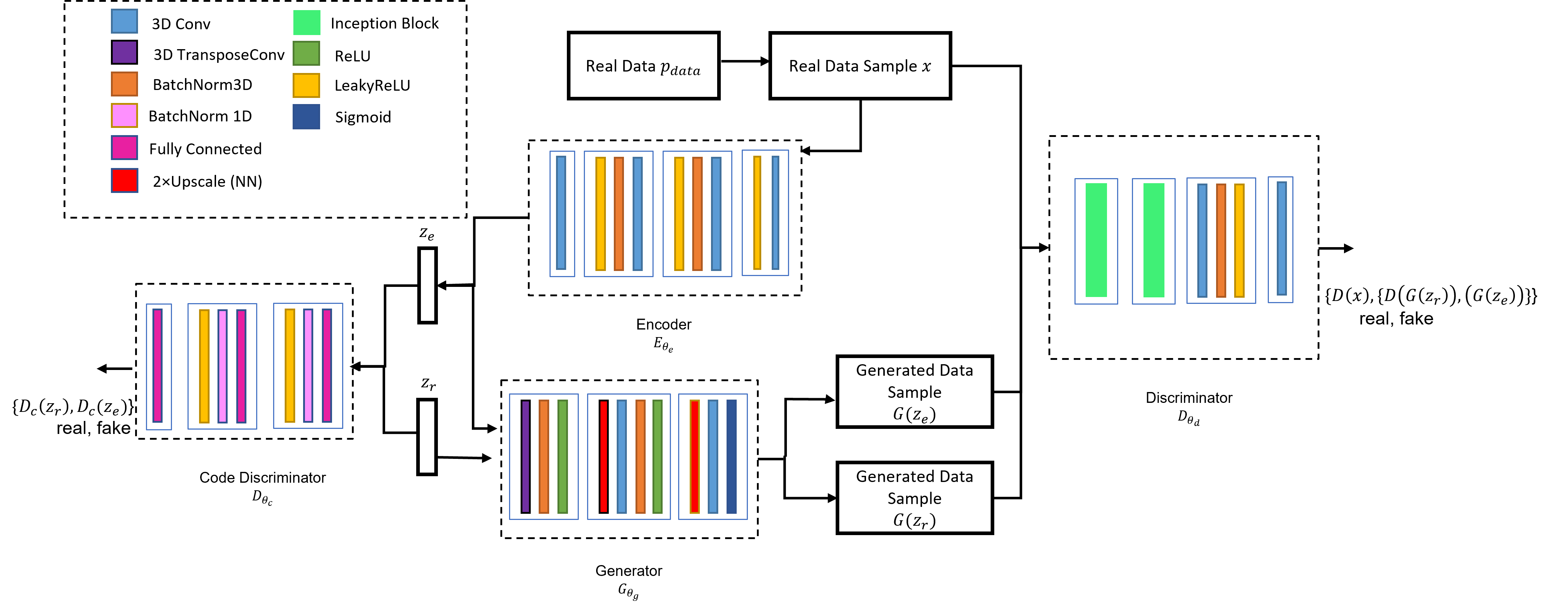}}
\hfill
\subfloat[Inception block structure {(Kernel sizes are shown)} \label{fig:inception_block}]{\includegraphics[width=0.65\textwidth]{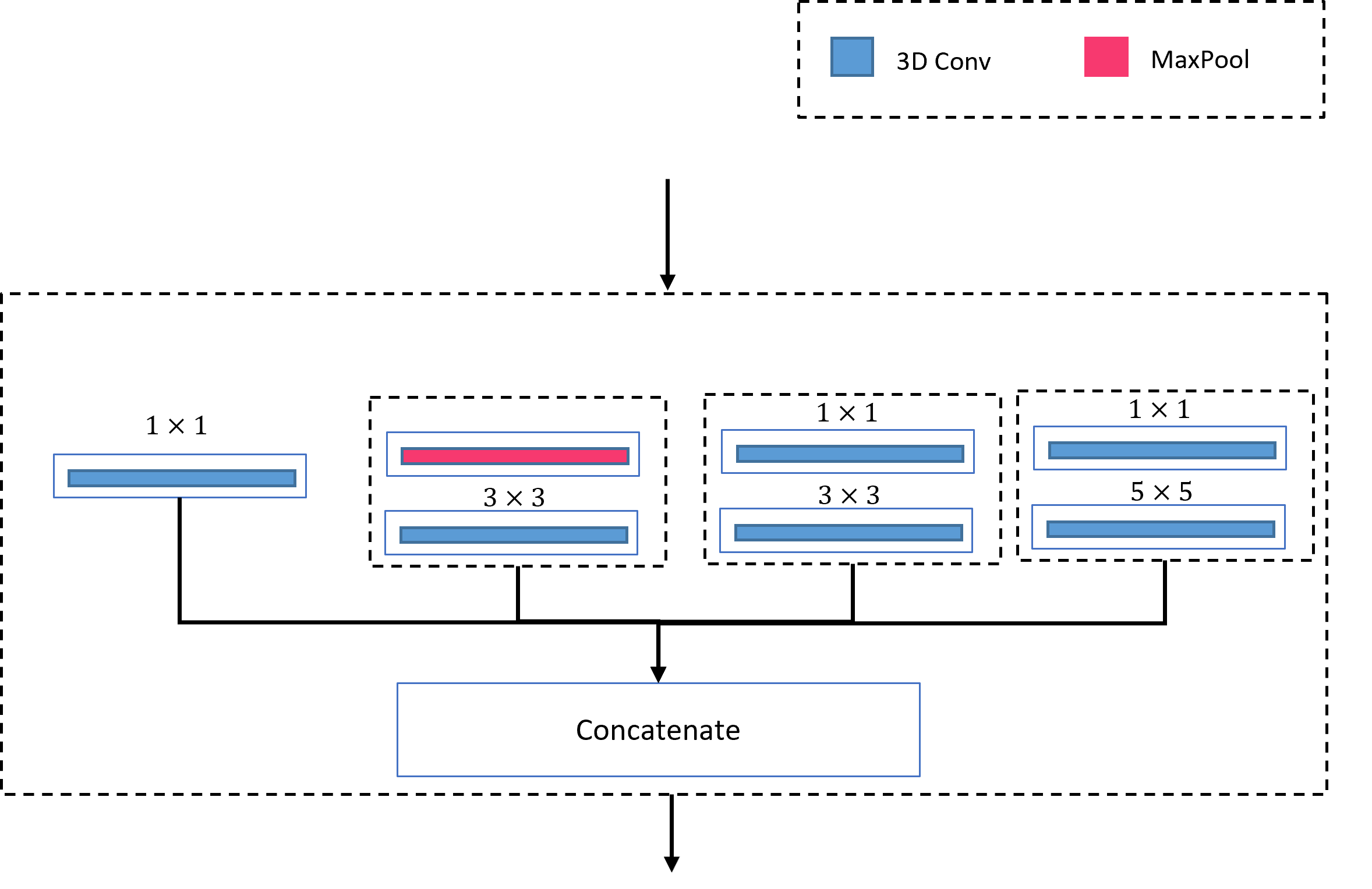}}
\caption{Improved 3D $\alpha$-GAN structure consists of four networks. The discriminator structure includes inception blocks.}
\label{fig:improved_Alpha_WGAN_with_networks}
\end{figure}

We design an improved version of 3D $\alpha$-GAN by altering the discriminator network's architecture.
Specifically, the new discriminator architecture illustrated in Figure~\ref{fig:improved_Alpha_WGAN_with_networks} incorporates an inception network structure \citep{szegedy2015going}. 
The inception block shown in Figure~\ref{fig:inception_block}, consists of four parallel convolutional blocks that affect the input. The output of convolutions are merged through channel dimension and passed to the next layers. Three of convolutional blocks include $1\times1$ kernels, which reduces the dimensionality of the channels.


Inception networks are designed to approximate and cover sparse local regions in a convolutional network. 
Therefore, it can help the network detect and extract connectivity features from the training data. 
The discriminator's capability in recognizing connectivity helps with the task of differentiation between real and fake data, and thus, penalizes the generated samples with low connectivity. 
Additionally, other networks' architectures are modified to generate the specific dimensions of our datasets (see Appendix~\ref{ap:model_arch}). 
The modified GAN structures are illustrated in Figure~\ref{fig:improved_Alpha_WGAN_with_networks_dis}. 

\subsection{Datasets}
We consider four distinct datasets in our numerical study, which are described below.

\begin{itemize}\setlength\itemsep{0.3em}
\item \emph{Synthetic connected 3D volumes}: This is a dataset of size 40,000 consisting of random spheres and ellipsoids with radius $(r_1,r_2,r_3)$ that are created in a $[16, 16, 16]$ 3D space. 
Voxel and mesh representation of a sample from the training set is illustrated in Figure~\ref{fig:connected_3D_volumes_voxel_mesh}.

\begin{figure}[!ht]
\centering
\includegraphics[width=0.8\textwidth]{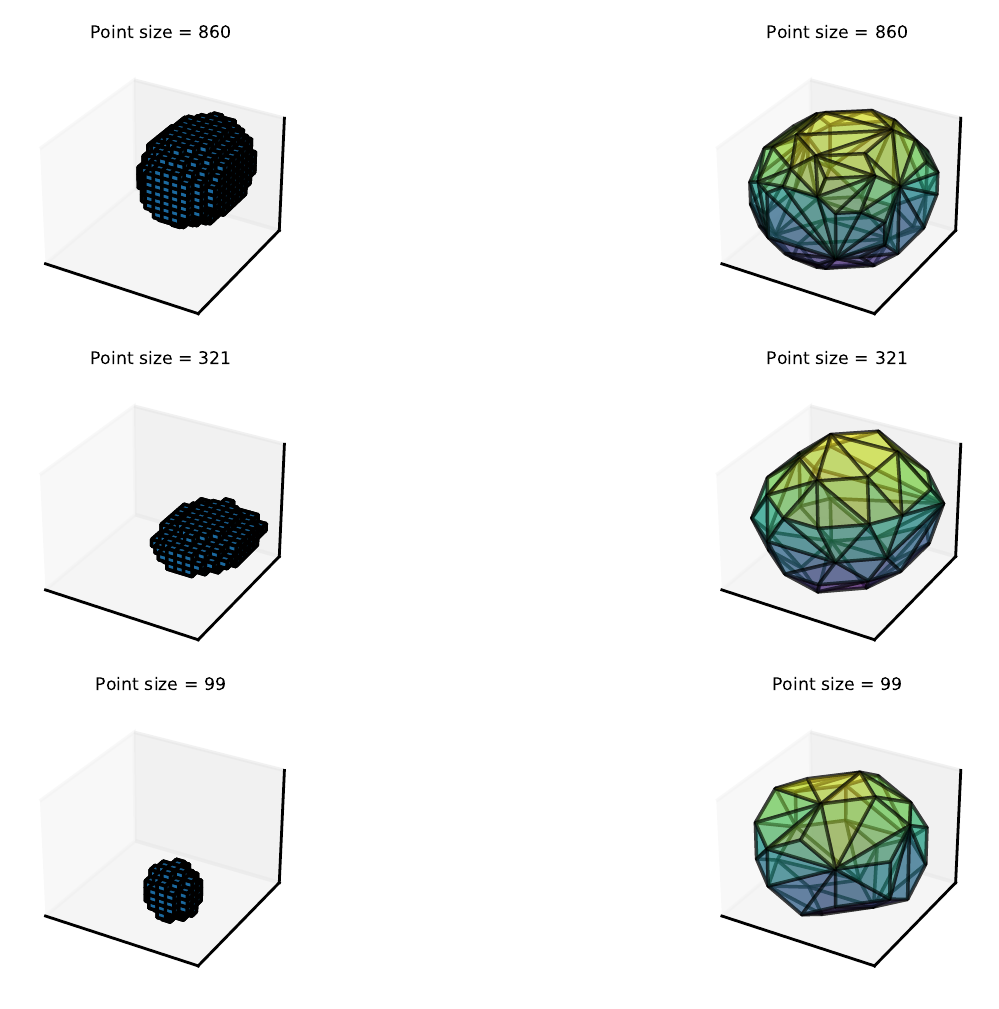}
\caption{Voxel and mesh representation of synthetic connected 3D volumes {(size of the connected volumes are shown in the title of images)}}
\label{fig:connected_3D_volumes_voxel_mesh}
\end{figure}




   \item \emph{Synthetic
   connected 3D volumes with packed spheres}: 
   Extending the previous dataset, a fixed number of smaller spheres are predefined inside each 3D volume. 
   However, since the sparse distribution of inside volumes can impact the learning performance, a denser version of the dataset is proposed in this part. 
   That is, the dataset from part one is modified, and, for seven predefined points that are distributed inside the volume, a random sphere with a radius of $r/4$ is created to cover a specific space in the generated original sample. We name these predefined points that are the center of subspheres as isocenters.
   A sample of the dataset is illustrated in Figure~\ref{fig:tumor_iso_sample_mesh_7}.
   Note that smaller spheres inside the main volume correspond to the isocenter locations in 3D tumor volume datasets.
   That is, the center of these smaller spheres can be treated as an individual isocenter.
   We also note that representing an isocenter as a smaller volume inside the main 3D volume is intuitive since the isocenters are the focal points of radiation delivery, and radiotherapy machines deliver dose focusing on this volume to eradicate tumor voxels in the vicinity of the isocenter, often forming a spherical or ellipsoidal area of impact~\citep{ghobadi2012automated}. A sample of a 3D tumor that is located between two healthy tissues or organs at risk (OARs) is illustrated in Figure~\ref{fig:matlab_tumor_iso_sample_cevik}. The six isocenter locations that are defined inside the tumor represent the focus points of transmitted radiations. 

\begin{figure}[!ht]
\centering
\subfloat[3D view of tumor with isocenter locations and OARs \label{fig:3D_matlab_isocenter}]{\includegraphics[width=0.47\textwidth]{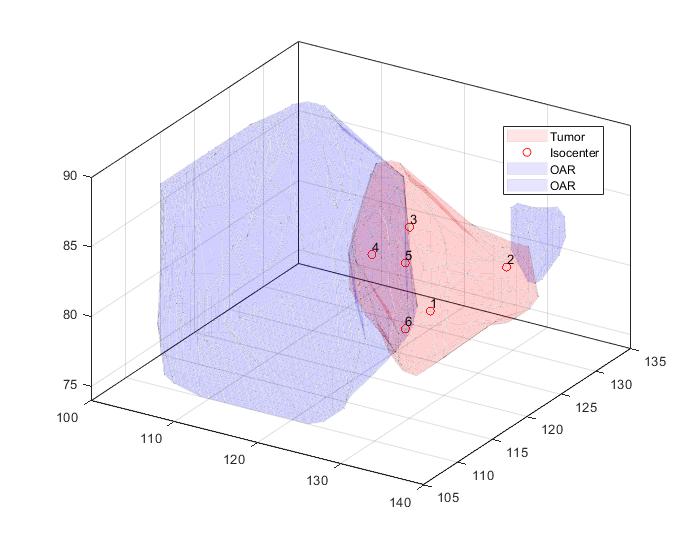}}
\hfill
\subfloat[X-Y view of tumor with isocenter locations and OARs    \label{fig:2D_matlab_isocenter}]{\includegraphics[width=0.47\textwidth]{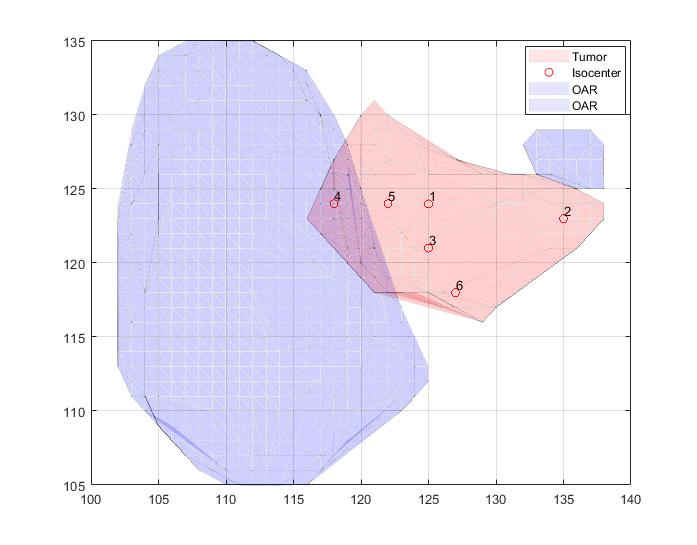}}

\caption{Sample tumor shape surrounded by two OARs and six defined isocenter locations}
\label{fig:matlab_tumor_iso_sample_cevik}
\end{figure}

    \begin{figure}[!ht]
    \centering
    \includegraphics[width=1\textwidth,height=5cm]{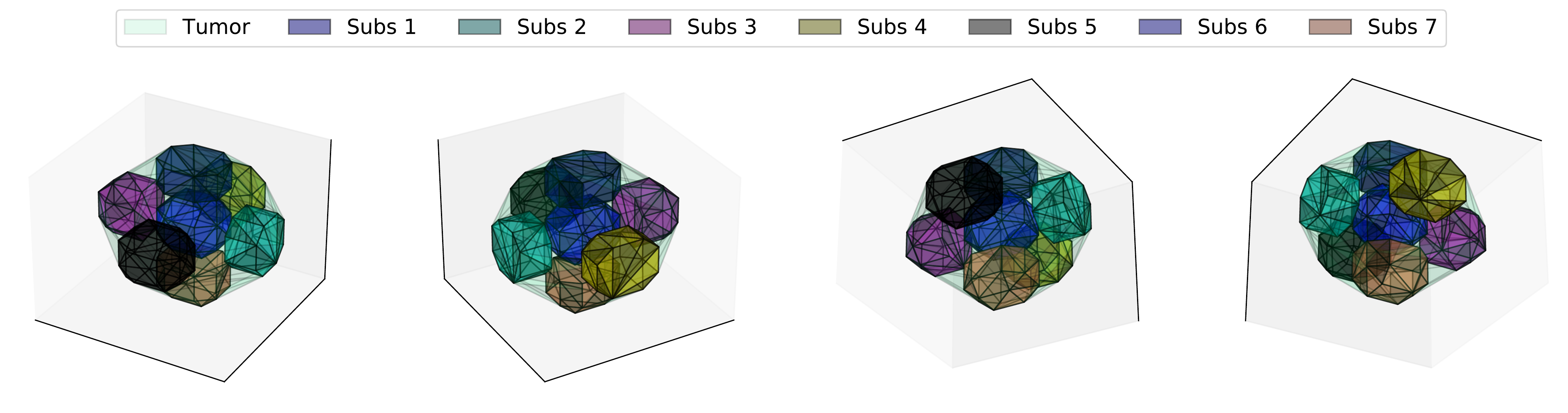}
    \caption{Mesh representation of synthetic connected volumes with filled smaller spheres illustrated from four different angels}
    \label{fig:tumor_iso_sample_mesh_7}
    \end{figure}



    \item \emph{Synthetic connected tumor volumes}:
    We generate a sample of connected volumes following the real distribution of 14 radiosurgery cases used in \citep{cevik2018modeling}. 
    Each voxel has a volume of 2$mm^3$, which in the space of $[16,16,16]$ represents the largest created tumor of size 12 $cm^3$.
    A sample of the dataset is illustrated in Figure~\ref{fig:tumor_gen_matlab_voxel_trisurf}.

    \begin{figure}[!ht]
        \centering
        \includegraphics[width=0.85\textwidth]{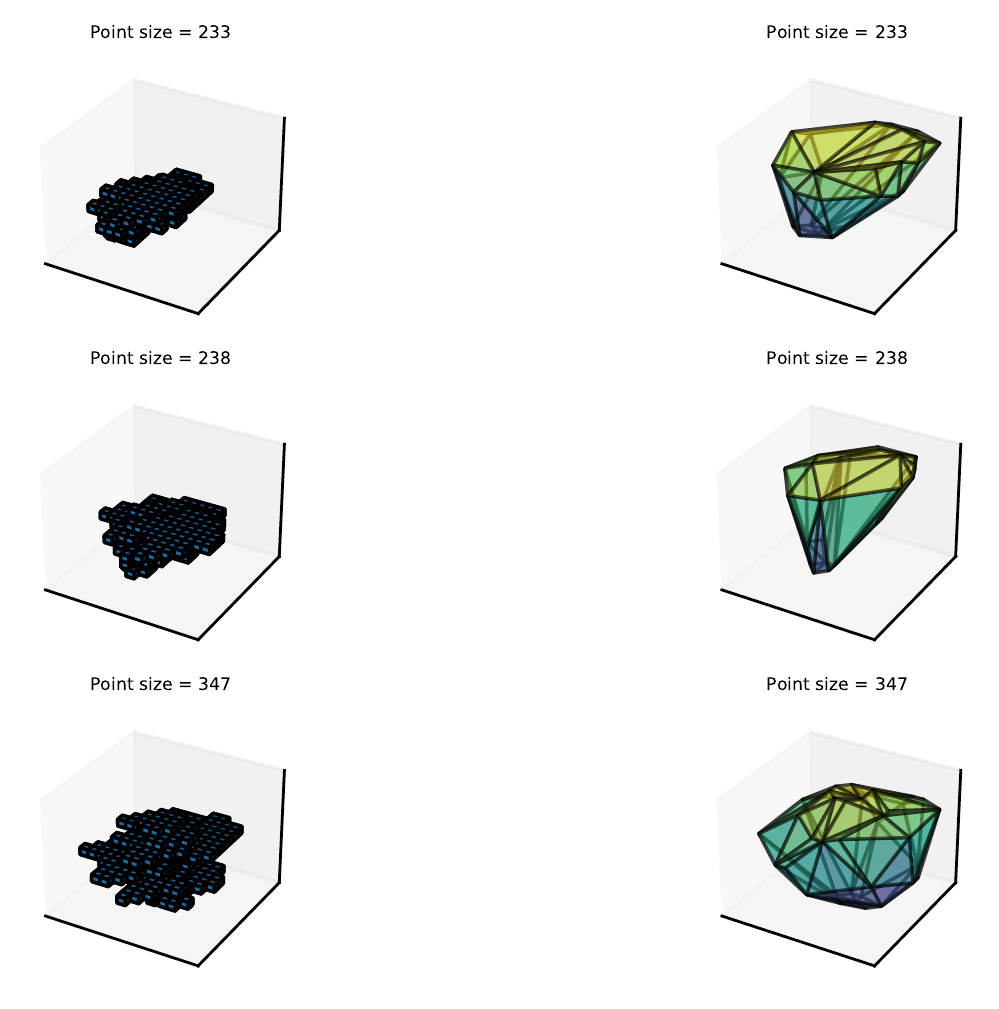}
        \caption{Voxel and mesh representation of training data for tumor volumes {(size of the connected volumes are shown in the title of images)}}
        \label{fig:tumor_gen_matlab_voxel_trisurf}
    \end{figure}
    

    \item \emph{Synthetic connected tumor volumes with packed spheres}:
    We use the synthetic tumor volume dataset as the basis of this dataset, and employ a hybrid of grassfire and sphere-packing algorithms proposed by \citep{ghobadi2012automated} to identify the center voxels of packed spheres (i.e., isocenters) inside the tumor volumes. We then return their covered volume's radius in the target tumors. A sphere-filling approach is used to surround each point with its covered radius.
    A sample of the dataset is illustrated in Figure~\ref{fig:tumor_iso_gen_matlab_trisurf}.
    \begin{figure}[!ht]
    \centering
    \subfloat[Sample connected volume with 11 subspheres \label{fig:tumor_iso_gen_matlab_trisurf2}]{\includegraphics[width=1\textwidth,height=4cm]{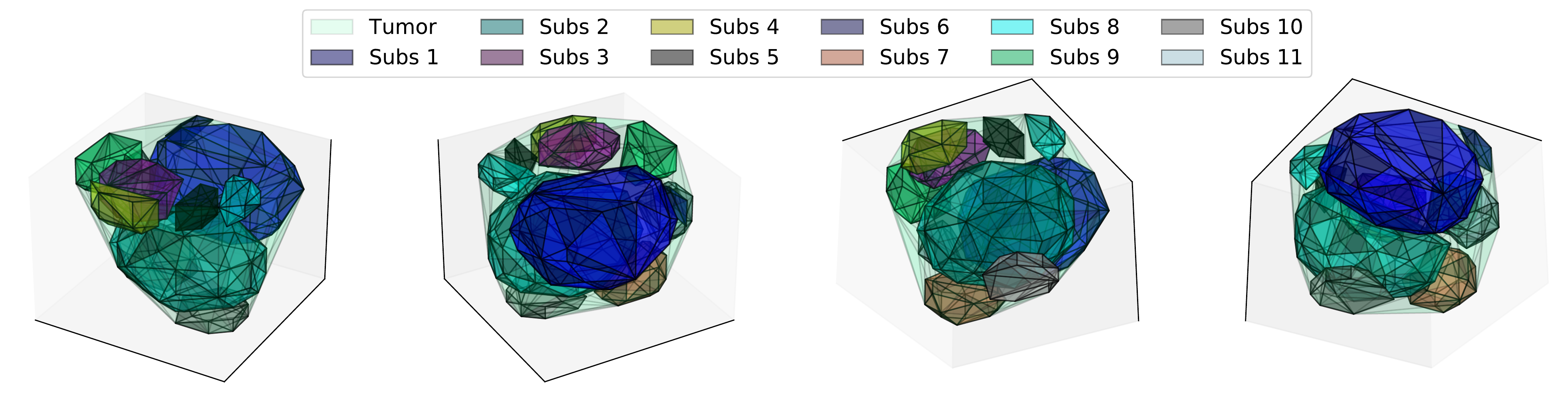}}
    \hfill
    \subfloat[Sample connected volume with 8 subspheres \label{fig:tumor_iso_gen_matlab_trisurf1}]{\includegraphics[width=1\textwidth,height=4cm]{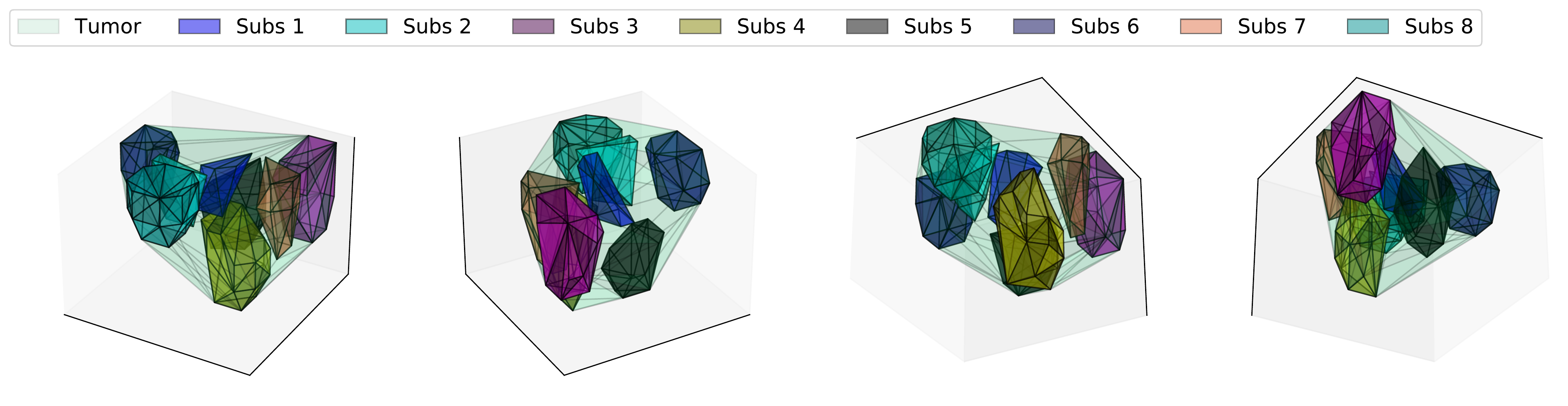}}
    \caption{Mesh representation of synthetic connected tumor volumes with filled subspheres from four different angles}
    \label{fig:tumor_iso_gen_matlab_trisurf}
    \end{figure}



\end{itemize}


\subsection{Experimental setup}
We evaluate our proposed 3D GAN architecture on four different datasets. 
We experiment with 3D convolutional GAN, WGAN-GP, and $\alpha$-GAN. 
PacGAN ~\citep{lin2018pacgan} approach is used in GAN and WGAN-GP architectures to avoid mode collapse. 
A detailed version of model structures is reported in Appendix~\ref{ap:model_arch}. 
Each model is trained on 40,000 training samples, and the evaluation metrics are reported over 10,000 generated samples by the generator network. 
For all experiments, Adam optimizer with a learning rate of 2e-4 is used. 
Models are trained for 500 epochs, and each experiment is reported five times to minimize the effect of random initialization. 
Since 3D space causes computational complications in our experiments, to avoid memory issues, a batch size of 40 is used. 
It is worth mentioning that the simple structure of our datasets makes it possible to use less complex CNN structures.
All the models were implemented using Python and PyTorch library, and experiments were performed on NVIDIA GeForce RTX 2070 GPU with 8 GB of GPU RAM.

\section{Numerical results}\label{sec:results}
In this section, we first present the definition and importance of performance metrics used to evaluate GAN architectures. We compare the performances of two baseline models including GAN and WGAN-GP with three different $\alpha$-GAN models over synthetic connected 3D volume datasets. 
Next, we choose the two best-performing models and resume the numerical comparison with connected tumor volume datasets. 
We note that the irregularities of tumor shapes compared to fully convex sphere/ellipsoid introduce a new challenge for the GAN models. 
We also provide samples of generated data to portray a clear visual understanding of the GANs' outputs.
Furthermore, to extend the discussion, we evaluate the models' performance on connected volumes packed with subspheres. The synthetic sphere/ellipsoids are assumed to be seven predefined filled spheres located at isocenters' specific coordinates.
On the other hand, the tumor data consists of a diverse number of subspheres with different coverage radiuses located at random coordinates. 
Therefore, the generation of synthetic connected volumes with packed spheres for sphere/ellipsoid is a less challenging task compared to the synthetic tumor data.

\subsection{Performance metrics}
Most popular evaluation metrics for GAN architectures are designed to assess image quality and diversity, as the majority of previous studies are focused on generating high-resolution detailed data such as face samples \citep{schonfeld2020u}. 
However, most of these metrics are not suitable for an effective evaluation of our methodology, because of our datasets' coarse details and the importance of pixel connectivity. 
Therefore, we propose novel metrics to test the generated samples. Proposed evaluation metrics can be divided into two categories, namely, shape metrics and location metrics. 
The shape metrics are used to evaluate the connectivity, convexity, and overall shape of the generated sample, whereas the location metrics are designed to evaluate the distribution of points and their relevant distances. 
More specifically, the location metrics are solely designed to evaluate the distribution of isocenters inside a connected volume. 

We note that both connected volume datasets have similar characteristics, and thus are evaluated using the same shape metrics. 
On the other hand, the connected volume datasets with packed spheres have differently designed sets of isocenters as the center of subspheres. 
The sphere/ellipsoid-shaped connected volumes have seven predefined isocenters, located at the center and in the middle of the main axis. 
We can extract the seven mentioned locations from the generated shapes, and compare their relative distances. 
On the other hand, tumor volumes' isocenters are located using the grassfire algorithm and can vary from 4 to 20 points.
Therefore, we exclude these metrics for the tumor volumes and only report the ratio of space that the generated subspheres cover inside the main volume.

Since there is no well-defined ideal value for the majority of our metrics, we report the KL divergence between the distributions of a metric for the training data vs the generated samples. 
We expect to get small KL divergence values to indicate the similarity of the two distributions, and how closely the generator followed the training data.
Performance metrics of our study related to shape evaluation are summarized as follows:

\begin{itemize}\setlength\itemsep{0.3em}

\item \emph{Connected volumes' sizes}: The size of the connected volumes presents the number of connected voxels that create the connected volume shape. The generated volumes' sizes distribution should ideally match the real data distribution. Therefore, we have another evaluation metric, where we report the KL divergence between the two distributions. 

\item \emph{Convexity ratio}:  We process the generated samples and find the best set of generated voxels that can generate a convex object. The number of generated voxels that are inside the convex shape divided by the total number of generated voxels defines the convexity ratio.
We note that only the sphere/ellipsoid connected volumes are fully convex, whereas the connected tumor volumes have a variety of shapes that can be convex or non-convex. 
Therefore, the convexity ratio for the former group is one, and for the latter group, it can range from 70\% to 100\%. 
Therefore, we solely report the KL divergence between the distribution of the convexity ratios.

\item \emph{Connectivity ratio}: 
Techniques on 3D shape generation typically involve post-processing the generated shapes and selecting the heavily clustered voxels as the final output. 
For this metric, we post-process the generated shapes by setting the average of points as the center and removing the points that have outlier distances to the center point. 
The ratio of the number of remaining points to the number of processed points is called the connectivity ratio. 
The ratio identifies how much post-processing is needed to reach a fully connected shape.

\item \emph{Coverage ratio}: We define certain thresholds for connectivity and convexity ratios by investigating the training data ratios. We then report the ratio of generated samples that meet the predefined thresholds of convexity and connectivity ratio as the coverage ratio. Ideally, all the generated shapes should be completely connected and the coverage ratio should be equal to one. 

\item \emph{3D moment invariants}: Moment invariants $(\Omega_1, \Omega_2, \Omega_3)$ are spatial descriptors used to quantify the distribution of an object's solid volume \citep{kazhdan2003rotation}. We use the distribution of 3D moments to evaluate the shape quality of the generated samples.


\item \emph{Shannon equitability index}: We use this index defined in Equation~\eqref{eq:shannon_formula} to measure the similarity (i.e., evenness) of different isocenters. 
For the case of sphere/ellipsoid connected volumes, since we designed the subspheres to be approximately similar, we can expect this measure to be near 1. 
However, the shape of subspheres are typically quite different in connected tumor volumes, and thus, we report the KL divergence of the Shannon index, that is,
\begin{align}
\label{eq:shannon_formula}
E_H = \frac{H}{\log(K)} \quad \text{where}\ H = -\sum_{i=1}^K p_i*\log(p_i)
\end{align}

\item \emph{Subspheres' coverage}: We measure the ratio of voxels that are covered by the packed spheres to the ratio of the entire connected volume. Ideally, we want the subspheres to cover the entire volume, and this coverage to be near 1.

\item \emph{Connected subspheres per connected volume:} We discussed the importance of the coverage ratio for volumes, and how it represented a connected object. For this measurement, since we defined each subsphere to be connected, we calculate the ratio of subspheres per volume that satisfy the coverage ratio. This measurement shows how introducing diversity in our images affects the model performance in generating multiple connected shapes. 
 
\end{itemize}

The following performance metrics are designed to evaluate the location and distribution of subspheres and isocenters. These are only applicable to the sphere/ellipsoid connected volumes, where the ideal isocenter locations are known and can be identified on a generated sample.
Proposed performance metrics related to distance evaluation are as follows:

\begin{itemize}\setlength\itemsep{0.3em}
      \item \emph{FD error}: Discrete Fréchet distance between the ground truth location of isocenters (processed on the output) and generated isocenters. 
      
      \item \emph{Ratio Mean Absolute Error}: MAE of the ratio between each isocenters' distance from the surface $D_s$ to the sum of the distance from the surface and center $D_c$ presented in Equation~\eqref{eq:ratio_formula} (excluding the isocenter located at the center). 
      Since we have located the isocenters in the middle of the main axis, we expect this ratio to be equal to 0.5. 
        \begin{align}\label{eq:ratio_formula}    
        \textit{MAE} &= \frac{1}{N}\sum_{i = 1}^{i = N}|0.5-\textit{Ratio}|\ \quad \text{where}\ \textit{Ratio} = \frac{D_s}{D_s + D_c} 
        \end{align}
        
        \item \emph{Target distance error}: For each generated shape the target distances between isocenters are generated, and the MAE error between generated and expected target distances is calculated using Equation~\eqref{eq:target_distances}. Target distances in the case of spheres are $\{r, r\sqrt{2}\}$ and in the case of ellipsoid are  $\{r_a$, $r_b$, $r_c$, $\frac{1}{2}\sqrt{(r_a)^2 + (r_b)^2}$,
         $\frac{1}{2}\sqrt{(r_a)^2 + (r_c)^2}$,
         $\frac{1}{2}\sqrt{(r_b)^2 + (r_c)^2}\}$.
        \begin{align}
        \begin{aligned}
        \textit{Target distance error} &=  \frac{1}{N}\sum_{j = 1}^{j = N}\sum_{i = 1}^{i = N}\min\limits_{i\neq j , i \neq \textit{center}}(\bigg |D(\textit{iso}[i], \textit{iso}[j]) \bigg |)
        \end{aligned}\label{eq:target_distances}    
        \end{align}

\end{itemize}

\begin{figure}[!ht]
\centering
\includegraphics[width=0.8\textwidth]{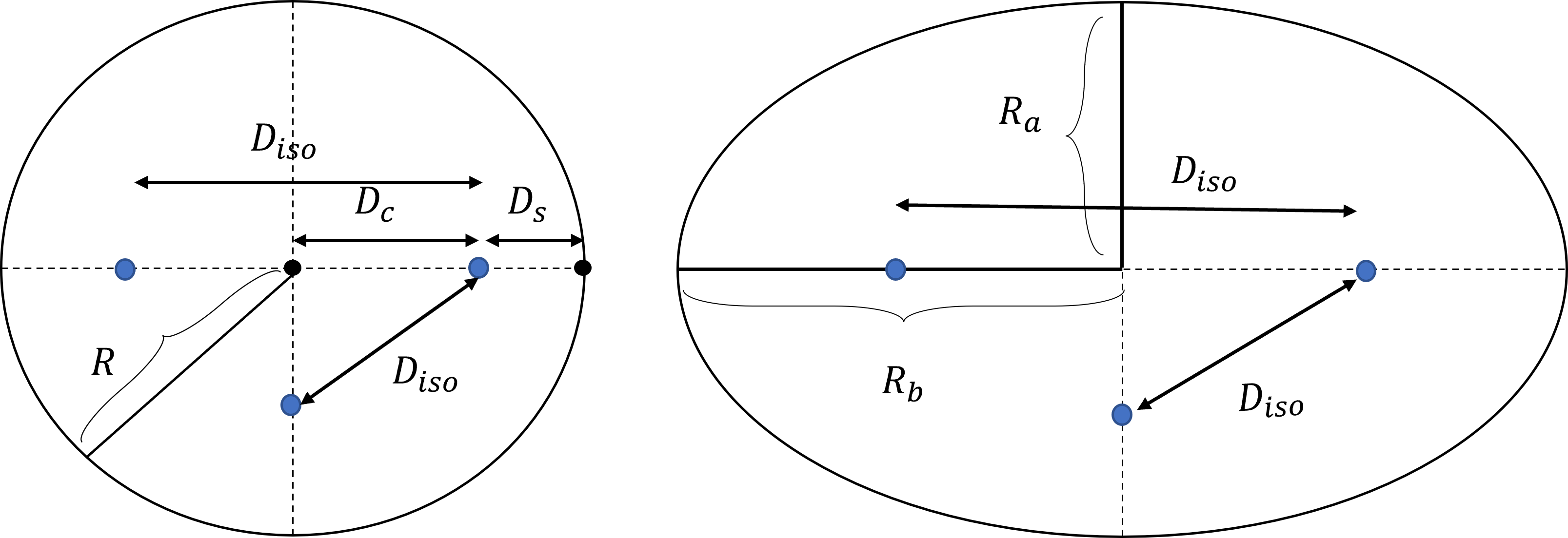}
\caption{2D shapes of generated samples and relative distances}
\label{fig:2d_distances}
\end{figure}

\subsection{Comparison of GAN performance over different datasets}
We report performance evaluation results for baseline GAN architectures and our proposed GAN over four synthetic datasets.
Firstly, we report our results for 3D connected volumes of sphere/ellipsoid shape. We elaborate on our findings and provide a sample of generated shapes. Next, we choose the best performing models and apply them to generated connected tumor volumes. Finally, we provide our results for connected volumes packed with subspheres.

\subsubsection{Results with connected 3D volumes}

\begin{figure}[!ht]
\centering
\subfloat[\footnotesize $\Omega_1$  (1.208, 0.784)\label{fig:OMEGA1_moment_data_40000}]{\includegraphics[width=0.33\textwidth]{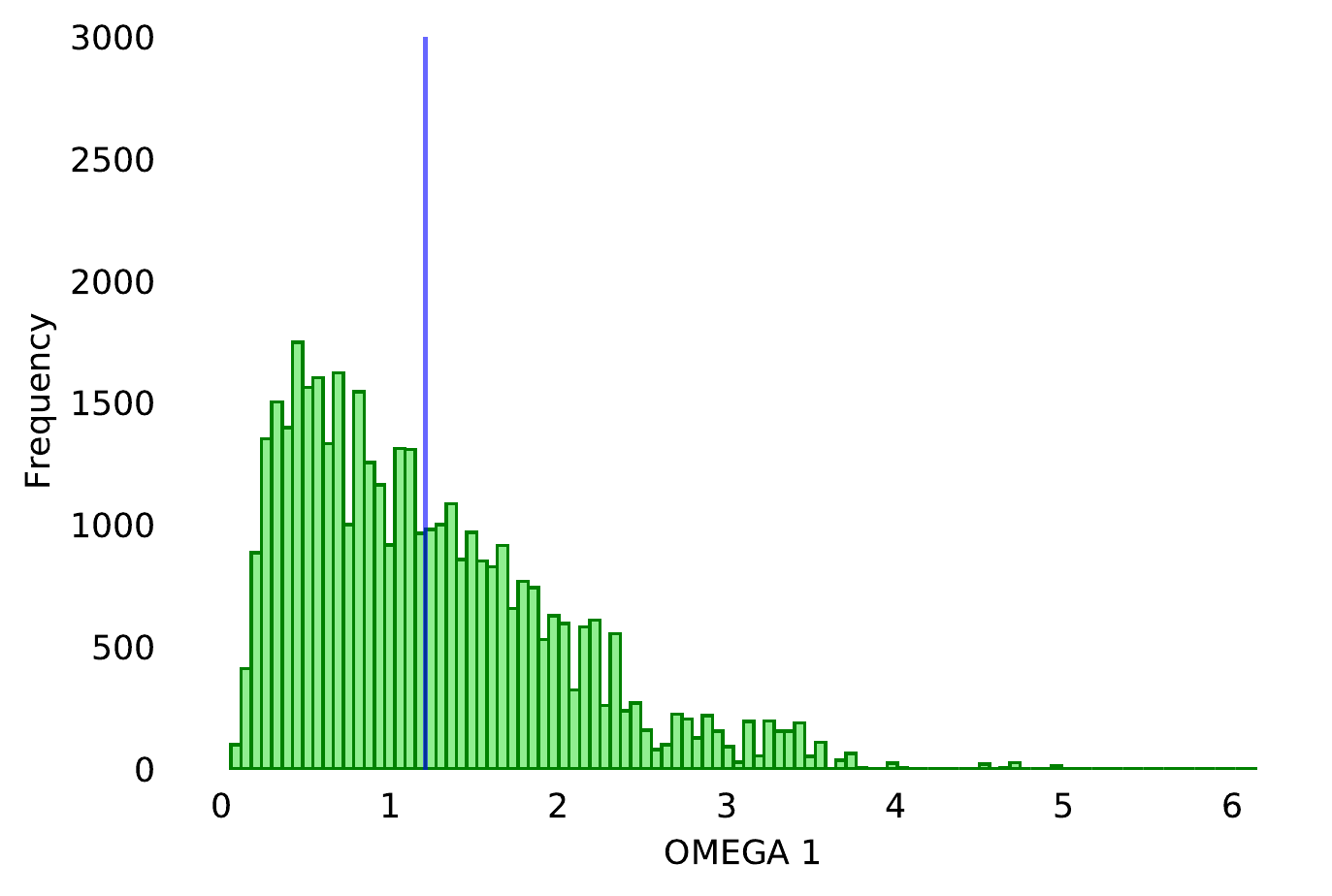}}
\hfill
\subfloat[\footnotesize $\Omega_2$  (7.784, 5.735)\label{fig:OMEGA2_moment_data_40000}]{\includegraphics[width=0.33\textwidth]{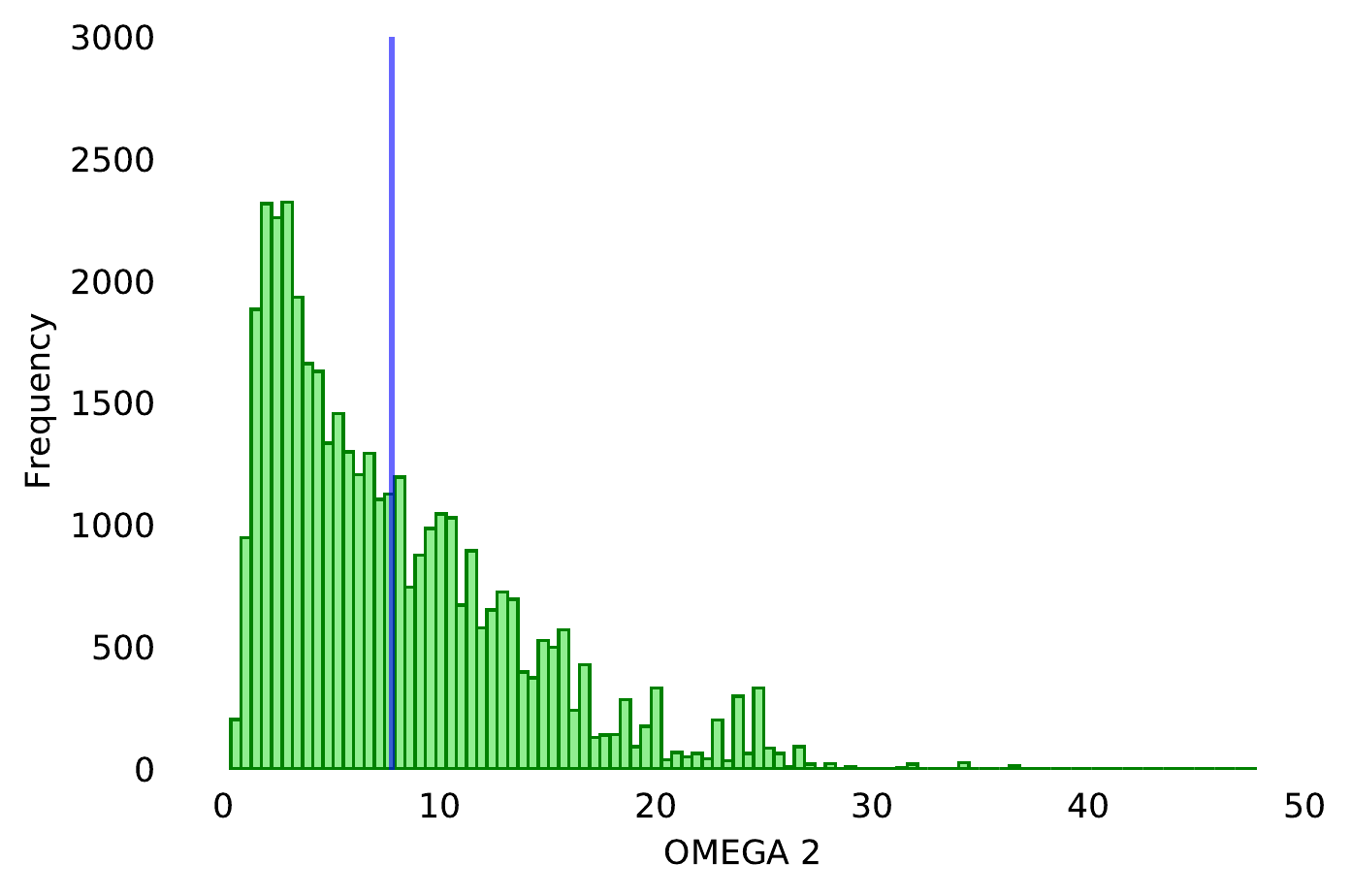}}
\hfill
\subfloat[\footnotesize $\Omega_3$  (69.552, 51.864)\label{fig:OMEGA3_moment_data_40000}]{\includegraphics[width=0.33\textwidth]{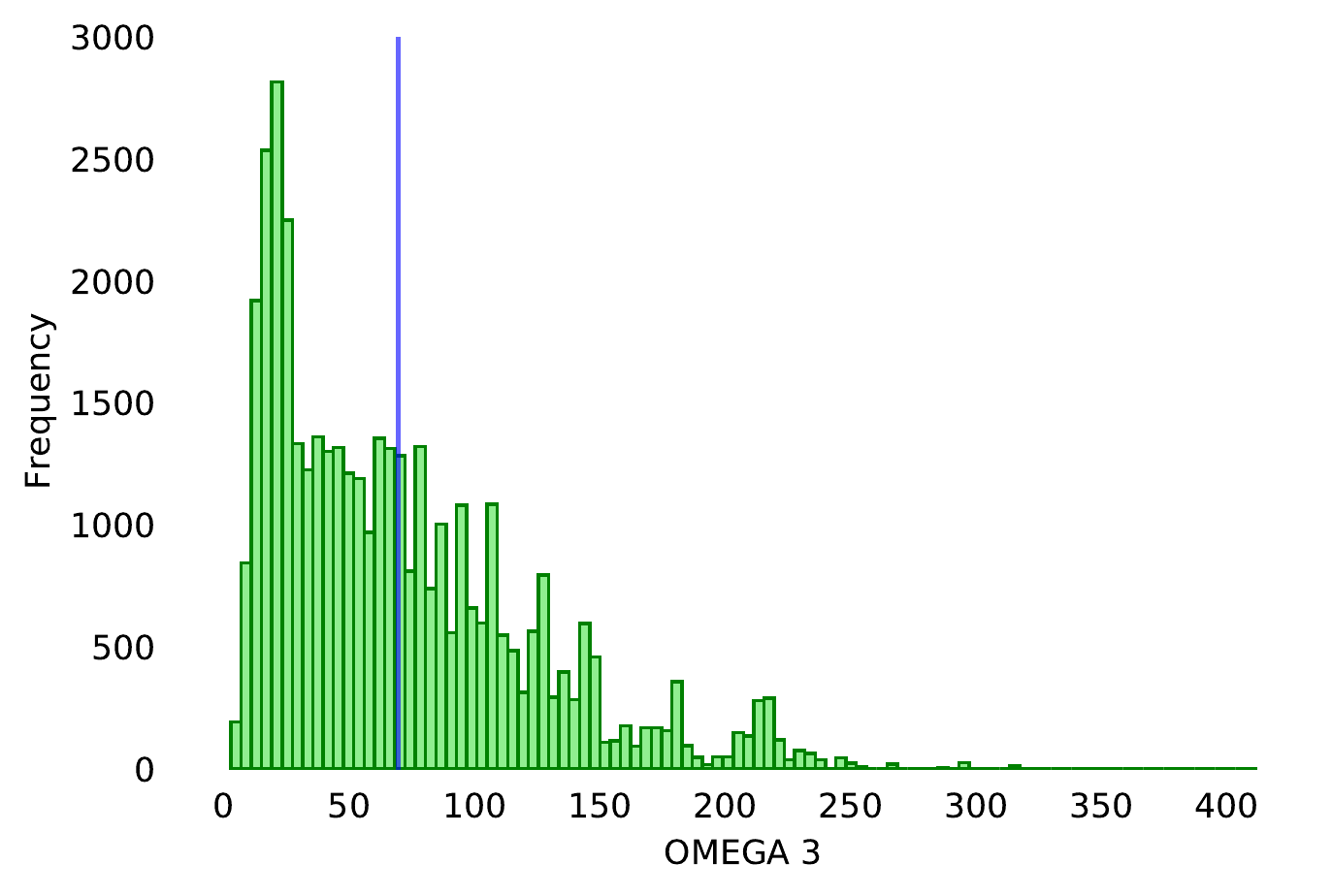}}
\caption{Moment invariant of the training data {(blue line represents the mean value)}}
\label{fig:moment_data_40000}
\end{figure}

We report the average of generated connected volumes' sizes, their coverage ratio, moment invariants, and KL divergence between the training data distribution of volumes' sizes, connectivity ratio, and convex ratio in Table~\ref{tbl:syntetic_results_tumor}. We also illustrated the distribution of 3D moment invariants and reported the mean and standard deviation for training data in Figure~\ref{fig:moment_data_40000}. 

\setlength{\tabcolsep}{3pt}
\renewcommand{\arraystretch}{1.8}
\begin{table}[!ht]
\caption{{\footnotesize Summary performance values for the synthetic connected 3D volume dataset evaluated for 10,000 samples averaged over 5 repeats }}
\label{tbl:syntetic_results_tumor}
\begin{center}
\resizebox{1.0\linewidth}{!}{
\begin{tabular}{lcccccccllll}
\toprule
& & & & & & & & \multicolumn{3}{c}{KL divergence} \\     
\cmidrule(lr){9-11} 
\mcps{\rrt GAN\\ model} & \mcps{\cnt Mode\\ collapse} & \mcps{\cnt Connected\\ volumes' sizes} & \mcps{\cnt Coverage\\ ratio} & $\Omega_1$ & $\Omega_2$ & $\Omega_3$ & & \mcps{\cnt Connected\\ volumes' sizes} & \mcps{\cnt Connectivity\\ ratio} & \mcps{\cnt Convexity\\ ratio}\\
\midrule
GAN & PacGAN & 579 $\pm$ 408 & 65 & 0.979 $\pm$ 0.606 &
4.822 $\pm$ 4.069 &
34.001 $\pm$ 32.889&& 
 0.200 $\pm$ 0.040&
 0.130 $\pm$ 0.039 &
  8.825 $\pm$ 0.489\\
WGAN-GP & PacGAN & 500 $\pm$ 346 & 92 & 1.166 $\pm$ 0.689& 7.058 $\pm$ 5.052& 59.273 $\pm$ 46.137  && 0.174 $\pm$ 0.036 & 0.077 $\pm$ 0.022 &  3.919 $\pm$ 0.192 \\
$\alpha$-GAN & -  & 681 $\pm$ 529 & 98 & 1.168 $\pm$ 0.764 & 7.439 $\pm$ 5.636 & 65.820 $\pm$ 51.385 && 0.158 $\pm$ 0.026 & 0.035 $\pm$ 0.010 & 0.105 $\pm$ 0.015\\ 
$\alpha$-GAN-CL$^\dagger$ & - & 709 $\pm$ 523 & 98 & 1.200 $\pm$ 0.766 & 7.605 $\pm$ 5.650 & 66.830 $\pm$  51.372 && 0.156 $\pm$ 0.026 & 0.034 $\pm$  0.010 & 0.153 $\pm$ 0.019\\  
$\alpha$-GAN++$^\ddagger$ & - & 748 $\pm$ 463 & 93 & 1.233 $\pm$ 0.702 & 
7.965 $\pm$ 5.216 &
70.799 $\pm$ 49.248 && 0.222 $\pm$ 0.048 & 0.045 $\pm$ 0.009 & 0.293 $\pm$ 0.032 \\
\bottomrule         
\end{tabular}
}
\\
{\tiny $^\dagger$: $\alpha$-GAN with connected loss,  $^\ddagger$: Improved $\alpha$-GAN}
\end{center}
\end{table}

In Table~\ref{tbl:syntetic_results_tumor}, we observe that the GAN model performs poorly compared to the rest of the models. This confirms the GAN's instability and \citet{kwon2019generation}'s assumption in replacing the simple GAN loss function with WGAN-GP.  
Other models show a high coverage ratio as well as similar 3D moment invariants to the training data. We can observe the effectiveness of $\alpha$-GAN by comparing it with WGAN. The coverage ratio is increased by 6\% and the convexity ratio is decreased from 3.919 to 0.105. The high coverage percentage shows the model's capability in generating connected shapes. We note that $\alpha$-GAN variants outperform the baseline GAN architectures by a significant margin. That is, the improved $\alpha$-GAN represents more similarity to the training data by comparing the values of moment invariants. However, the other two variants outperform improved $\alpha$-GA in terms of KL divergence values for different evaluation metrics. This might be because the high convexity ratios of sphere/ellipsoid shapes deteriorate the models' performance.


\begin{figure}[!ht]
\centering
\subfloat[\footnotesize $\Omega_1$  (1.242, 0.708) \label{fig:OMEGA1_moment_data_sample}]{\includegraphics[width=0.33\textwidth]{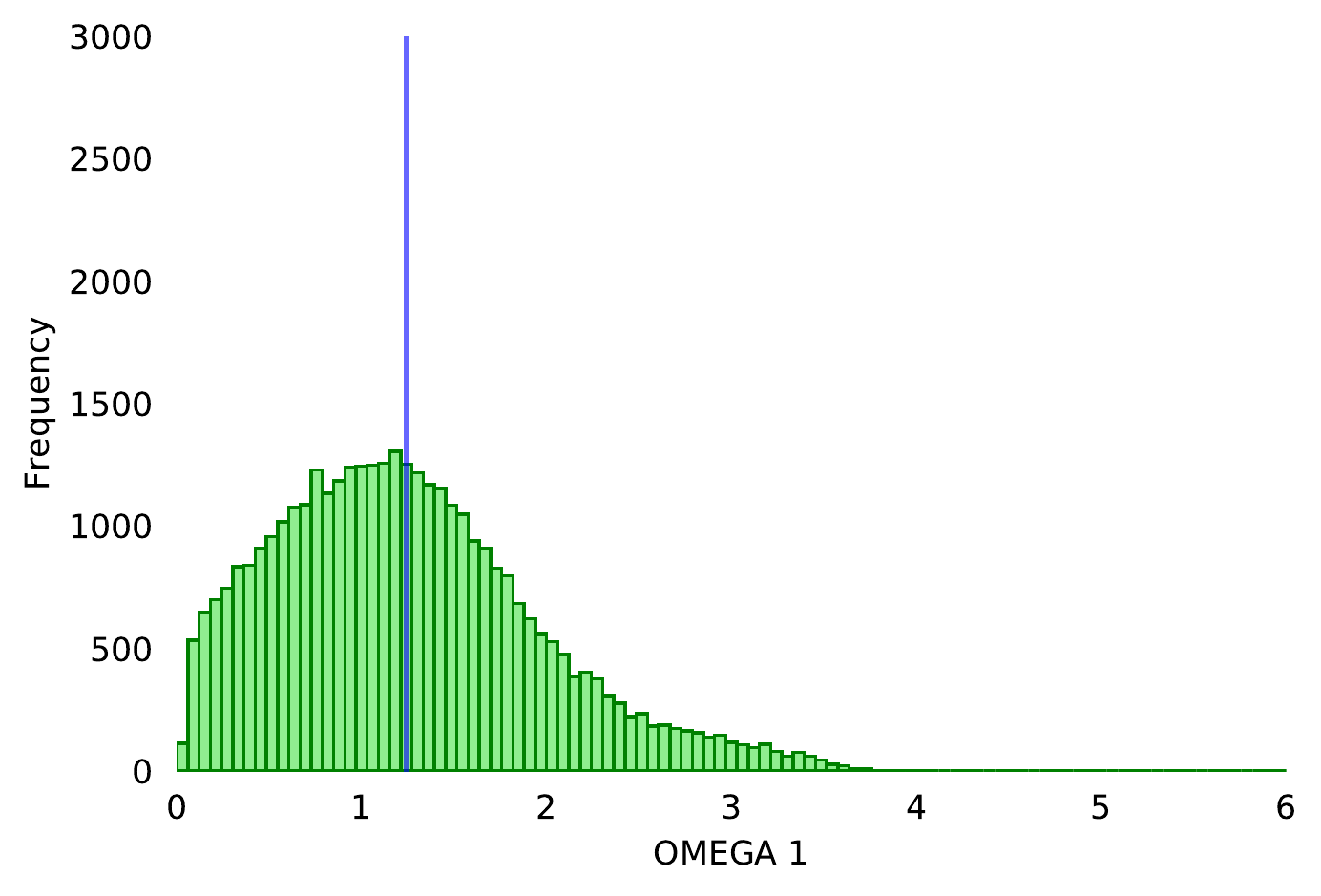}}
\hfill
\subfloat[\footnotesize $\Omega_2$ (8.019, 5.269) \label{fig:OMEGA2_moment_data_sample}]{\includegraphics[width=0.33\textwidth]{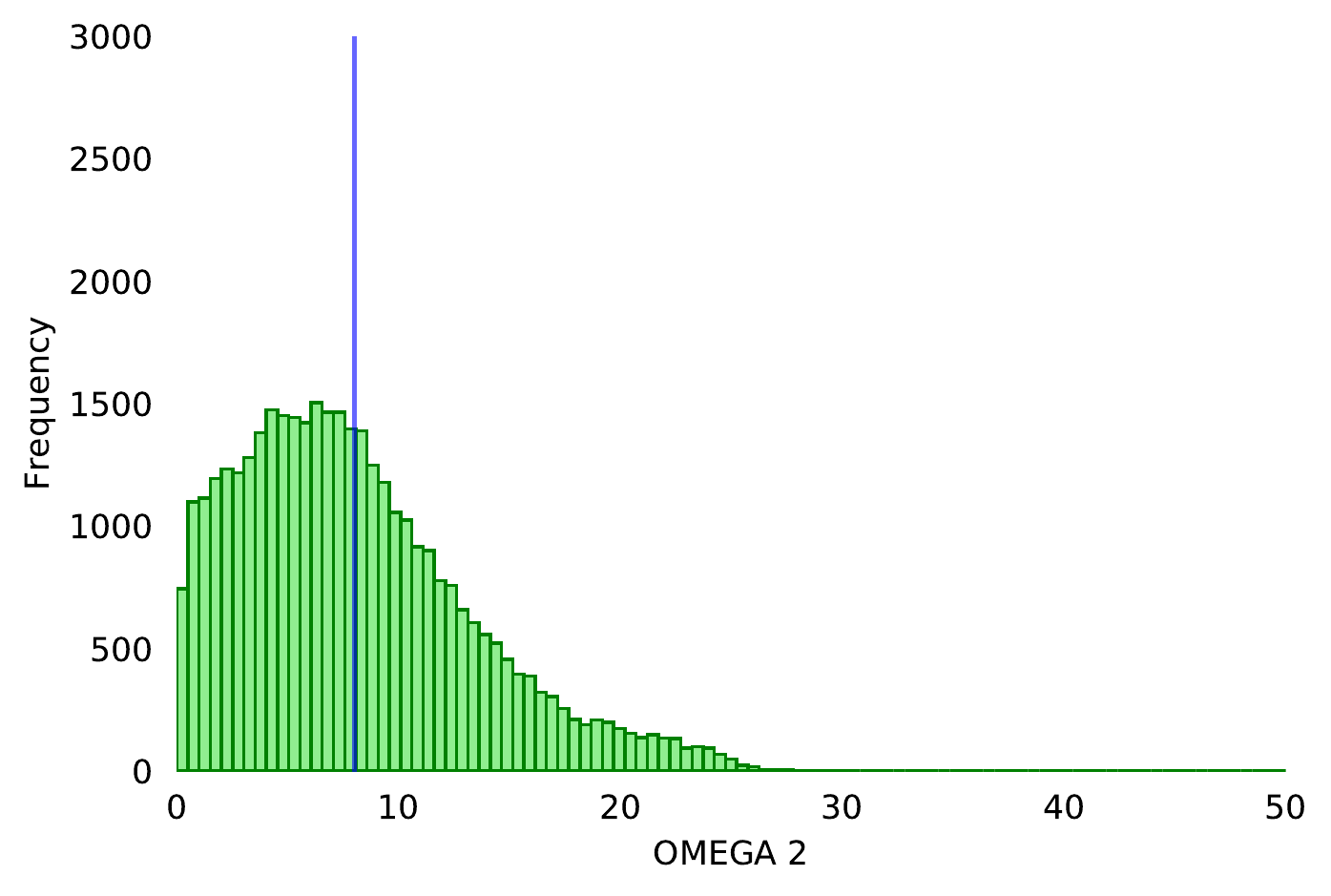}}
\hfill
\subfloat[\footnotesize $\Omega_3$ (71.269, 49.841) \label{fig:OMEGA3_moment_data_sample}]{\includegraphics[width=0.33\textwidth]{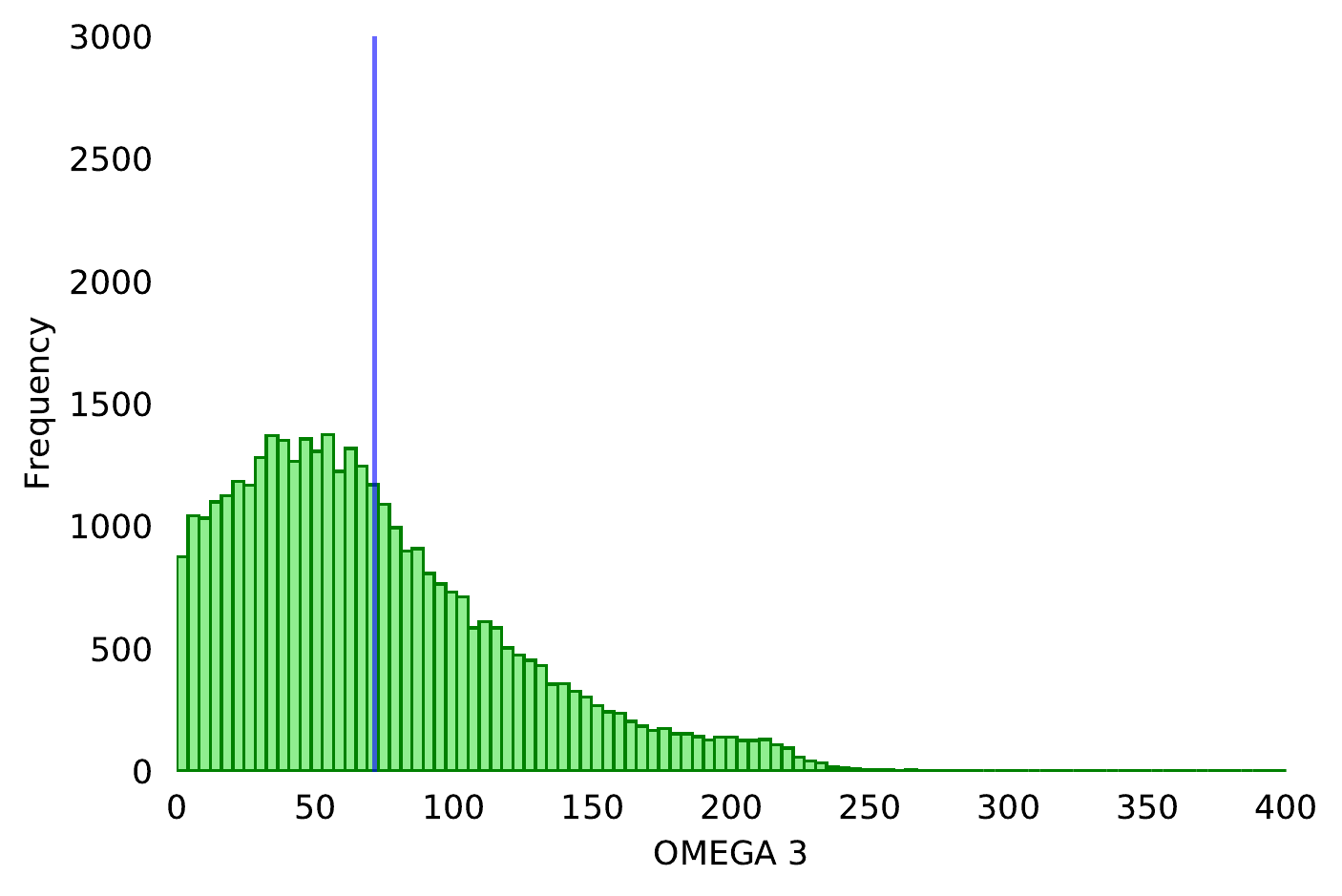}}
\caption{Moment invariants of generated data for the improved $\alpha$-GAN over 40000 samples {(blue line represents the mean value)}}
\label{fig:moment_data_sample}
\end{figure}
\begin{figure}[!ht]
    \centering
    \includegraphics[width=0.85\textwidth]{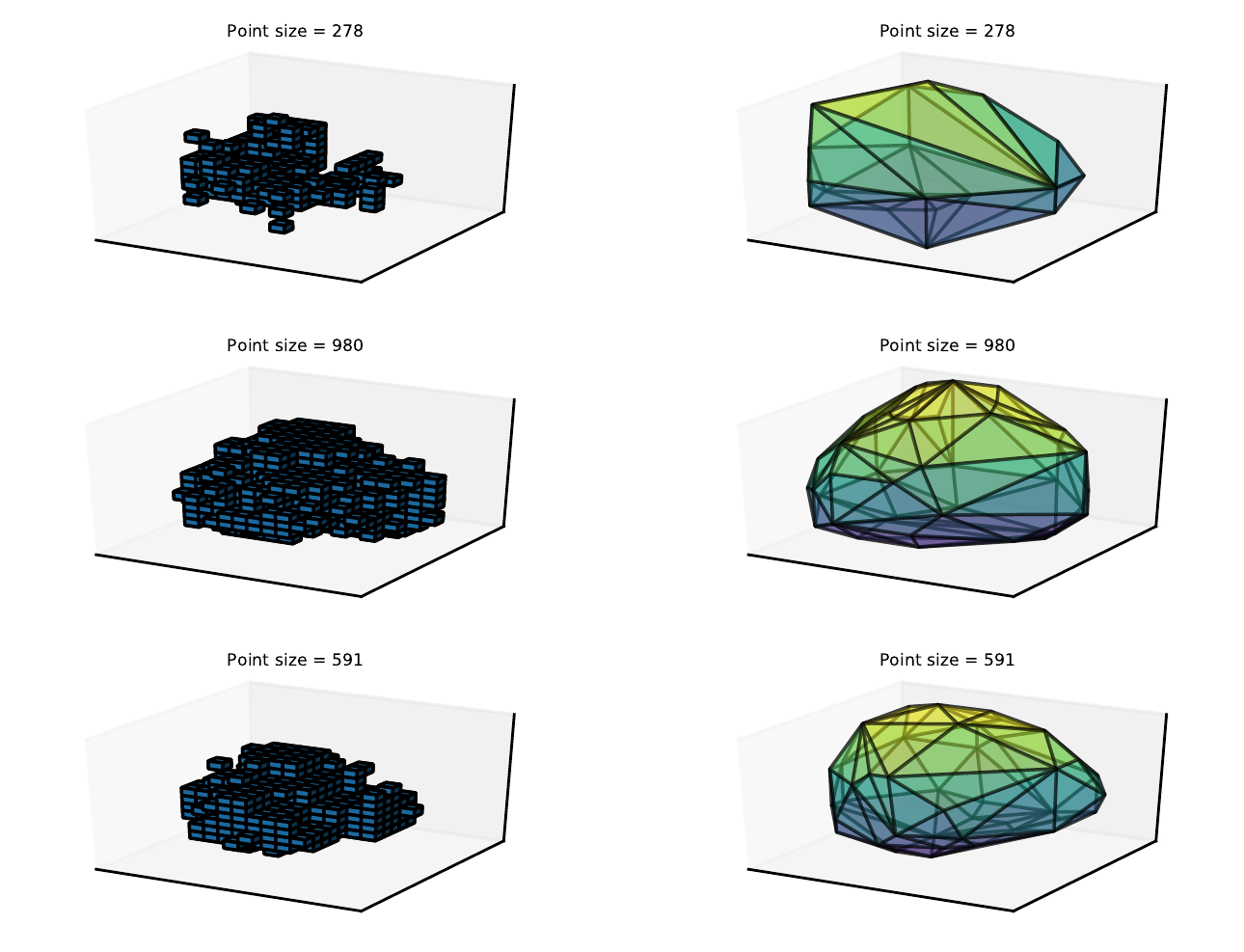}
    \caption{Voxel and mesh representation of sample generated connected volumes' shapes for the improved $\alpha$-GAN {(sizes of the shapes are shown in the title of images)}}
    \label{fig:tumor_gen_sample_trisurf}
\end{figure}

We illustrate the 3D moment invariants of generated samples in Figure~\ref{fig:moment_data_sample} to compare those with training data. The mean and standard deviation values show that the generated data consistently exhibit a higher mean and standard deviation.
A sample of generated connected volumes' shapes is illustrated in  Figure~\ref{fig:tumor_gen_sample_trisurf}. All three shapes are highly convex and fully connected.

\begin{figure}[!ht]
\centering
\subfloat[\footnotesize $\Omega_1$  (0.887, 0.154)\label{fig:OMEGA1_moment_data_tumor_Matlab_40000}]{\includegraphics[width=0.33\textwidth]{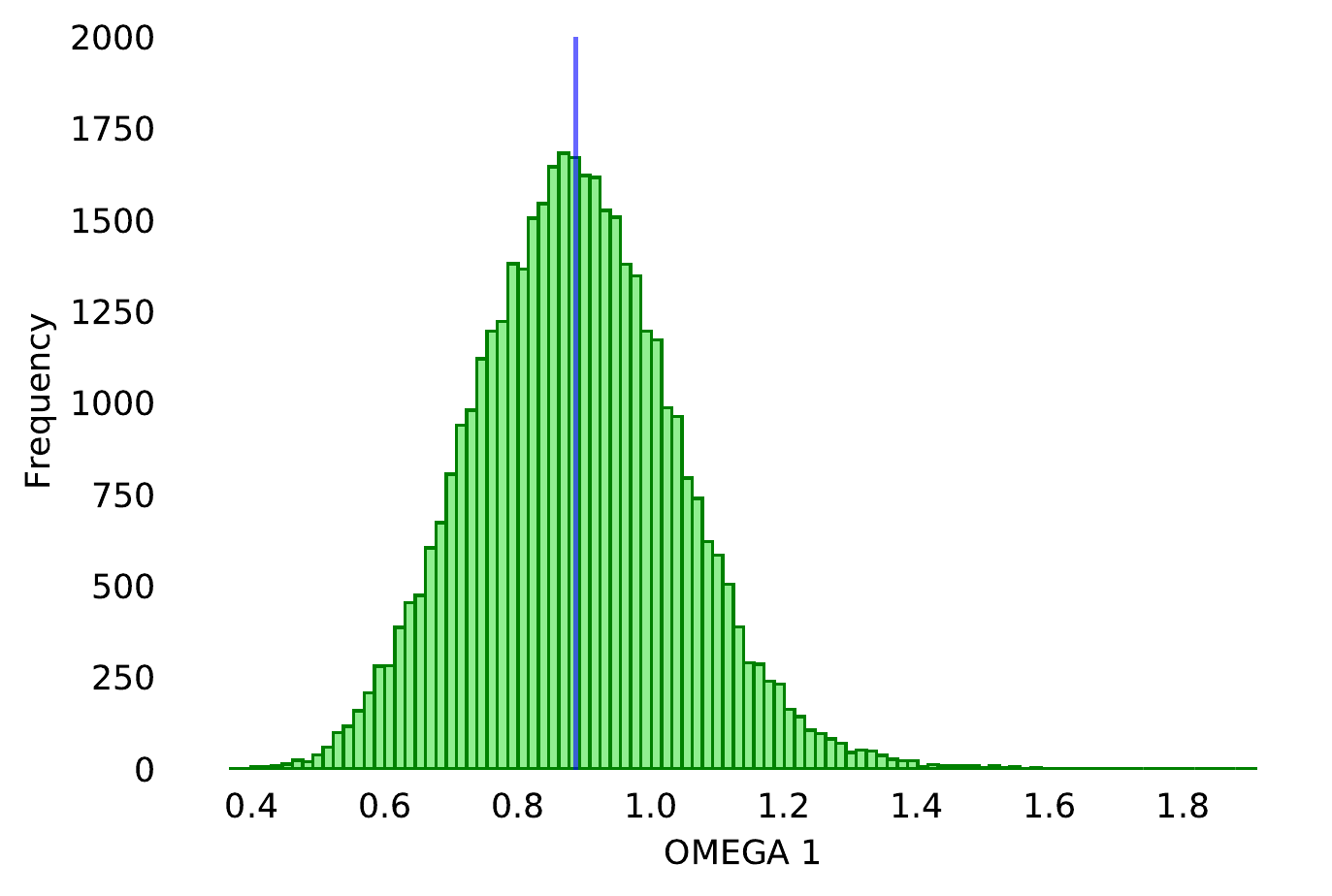}}
\hfill
\subfloat[\footnotesize $\Omega_2$  (3.991, 1.139)\label{fig:OMEGA2moment_data_tumor_Matlab_40000}]{\includegraphics[width=0.33\textwidth]{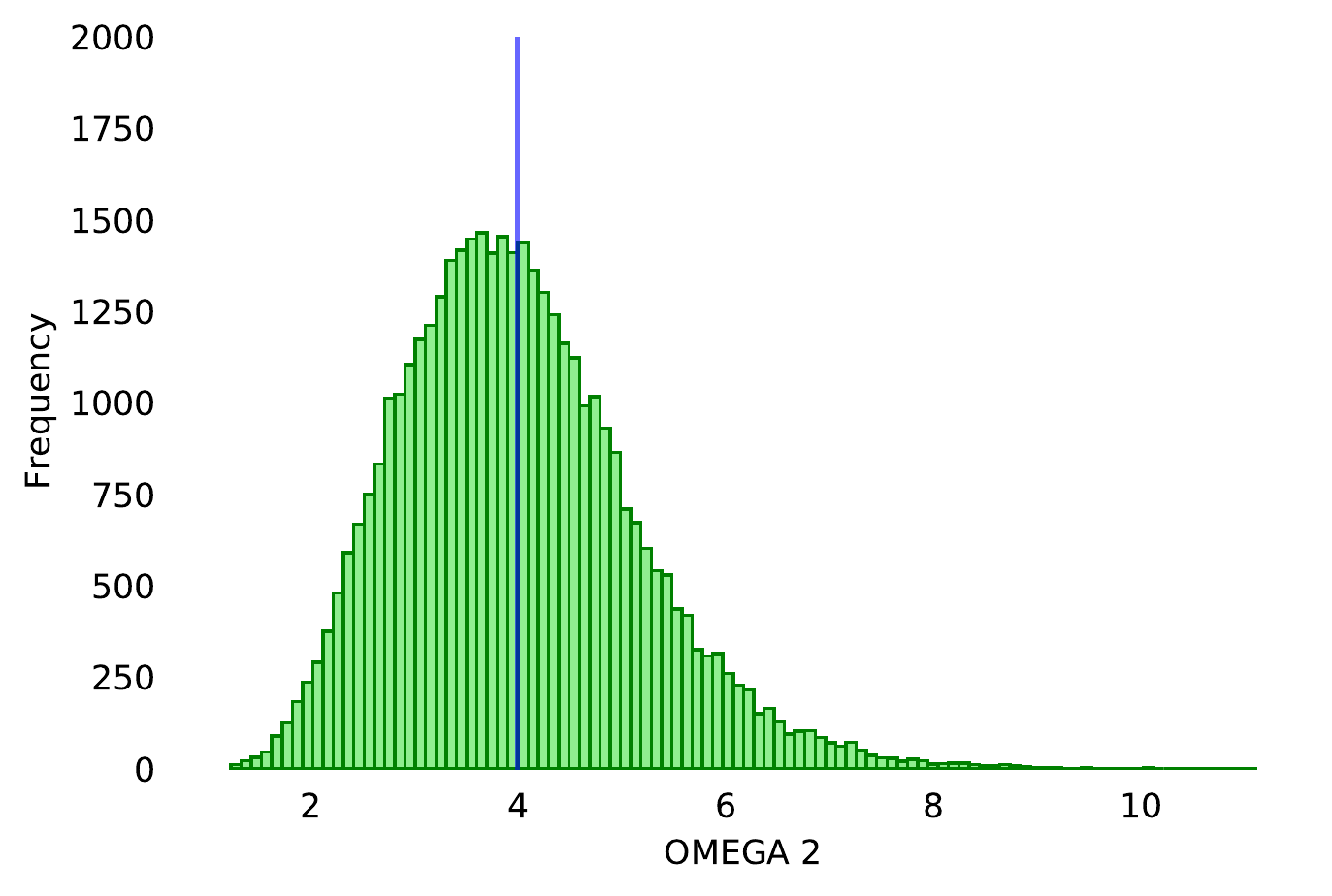}}
\hfill
\subfloat[\footnotesize $\Omega_3$  (26.761, 11.304)\label{fig:OMEGA3moment_data_tumor_Matlab_40000}]{\includegraphics[width=0.33\textwidth]{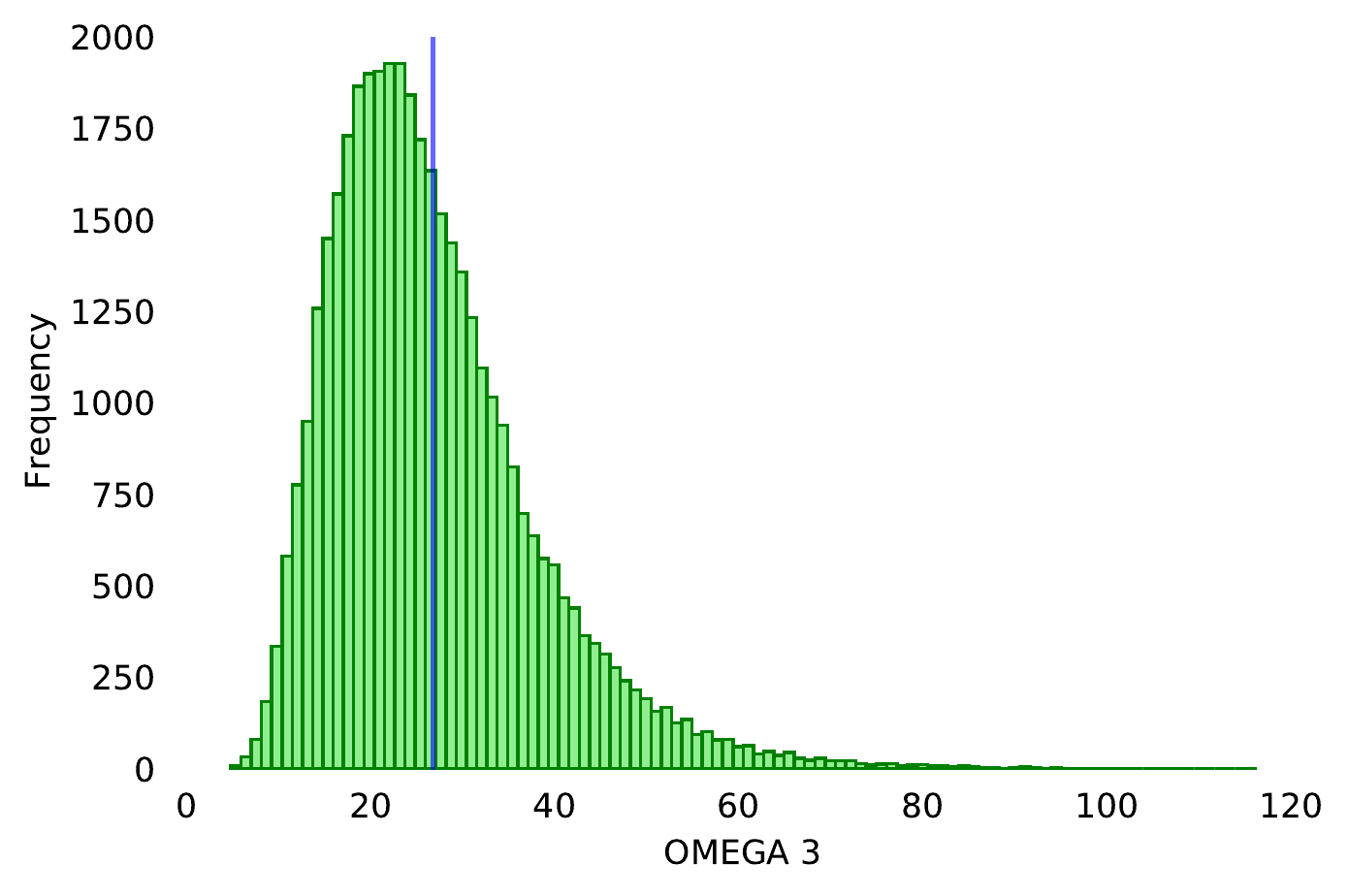}}
\caption{Moment invariant of the training data {(blue line represents the mean value)}}
\label{fig:moment_data_tumor_Matlab_40000}
\end{figure}

The connected 3D tumor volumes dataset consists of fully connected volumes with more variable shapes, as they can reflect convex or non-convex characteristics in different parts of the shape. We illustrate the distribution of the 3D moment invariants $(\Omega_1, \Omega_2, \Omega_3)$ in Figure~\ref{fig:moment_data_tumor_Matlab_40000}. A comparison with Figure~\ref{fig:moment_data_40000} for sphere/ellipsoid connected volumes portrays the difference in shapes. 

The results for the best-performing models are reported in Table~\ref{tbl:results_tumor_matlab}. The high coverage ratio, low KL divergence, and close 3D moment invariants corroborate our previous findings. However, we note that the improved $\alpha$-GAN loses its advantage in generating similar moment invariant values for all three moments. 

\setlength{\tabcolsep}{4pt}
\renewcommand{\arraystretch}{1.8}
\begin{table}[!ht]
\caption{{\footnotesize Summary performance values for the synthetic connected 3D tumor volume dataset evaluated for 10,000 samples averaged over 5 repeats }}
\label{tbl:results_tumor_matlab}
\begin{center}
\resizebox{1.0\linewidth}{!}{
\begin{tabular}{lccccccllll}
\toprule
& & & & & & & \multicolumn{3}{c}{KL divergence} \\     
\cmidrule(lr){8-10} 
\mcps{\rrt GAN\\ model} & \mcps{\cnt Connected\\ volumes' sizes} & \mcps{\cnt Coverage\\ ratio} & $\Omega_1$ & $\Omega_2$ & $\Omega_3$ & & \mcps{\cnt Connected\\ volumes' sizes} & \mcps{\cnt Connectivity\\ ratio} & \mcps{\cnt Convexity\\ ratio}\\
\midrule

$\alpha$-GAN &272 $\pm$  77 & 96 & 0.945 $\pm$ 0.171 & 4.044 $\pm$ 1.209 &  25.612 $\pm$ 10.787 && 1.728 $\pm$ 0.193 & 0.002 $\pm$ 0.000 & 0.045 $\pm$ 0.011\\
$\alpha$-GAN-CL$^\dagger$ & 269 $\pm$ 92 & 93 &  0.878 $\pm$ 0.160 & 3.508 $\pm$ 1.096 &
20.373  $\pm$ 9.235 & & 2.145 $\pm$ 0.203 & 0.004 $\pm$ 0.000 & 0.069 $\pm$0.017\\  
$\alpha$-GAN++$^\ddagger$ & 332 $\pm$ 103 & 89 & 0.957 $\pm$ 0.166 & 4.120 $\pm$ 1.348 &  26.260 $\pm$ 12.035 && 2.744 $\pm$ 0.261 & 0.011 $\pm$ 0.002 & 0.056 $\pm$  0.014 \\
\bottomrule         
\end{tabular}
}
{\tiny $^\dagger$: $\alpha$-GAN with connected loss,  $^\ddagger$: Improved $\alpha$-GAN}
\end{center}
\end{table}

\begin{figure}[!ht]
\centering
\subfloat[\footnotesize $\Omega_1$  (0.992, 0.230)\label{fig:OMEGA1_moment_data_tumor_Matlab}]{\includegraphics[width=0.33\textwidth]{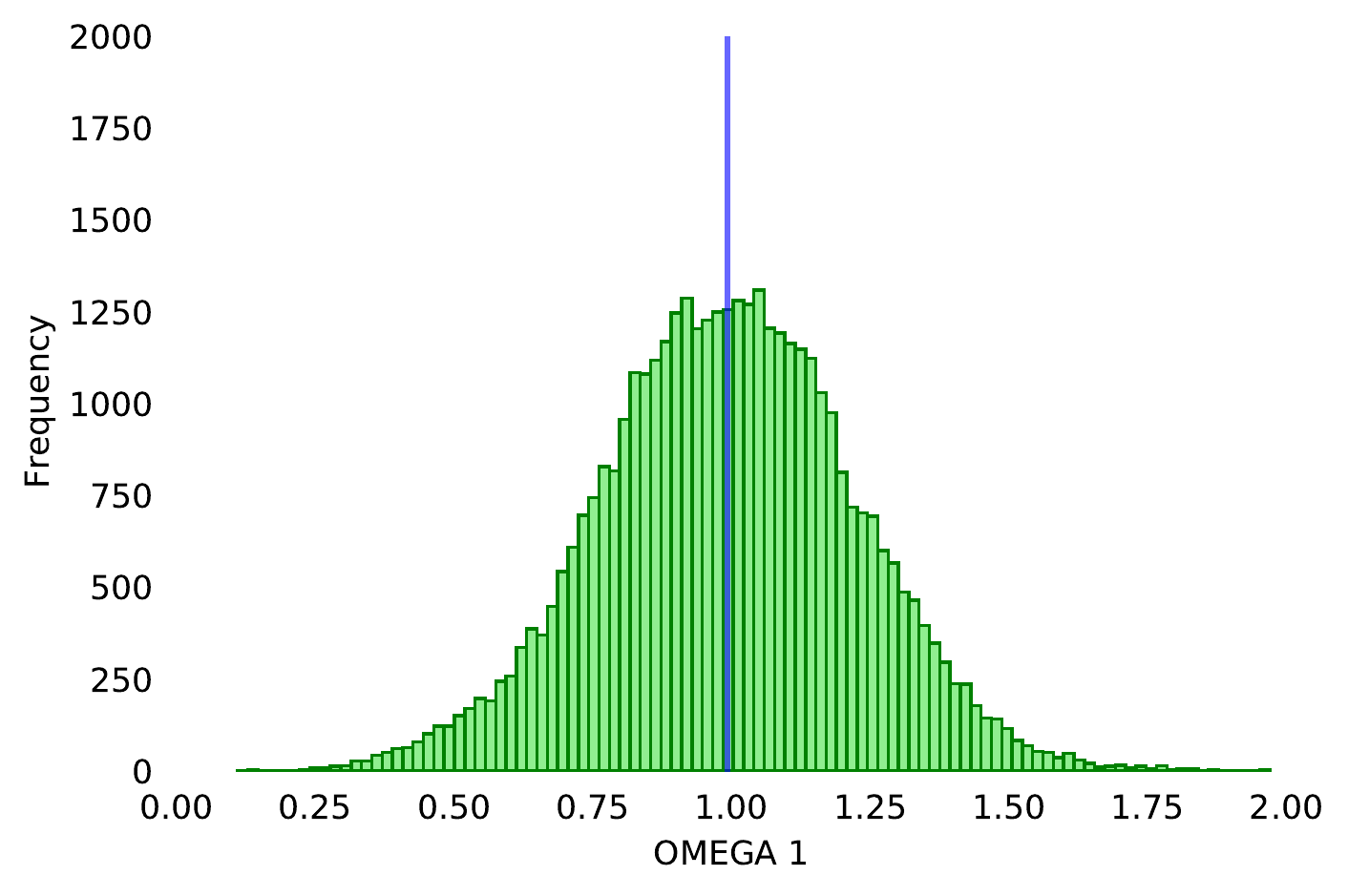}}
\hfill
\subfloat[\footnotesize $\Omega_2$  (4.129, 1.576)\label{fig:OMEGA2_moment_data_tumor_Matlab}]{\includegraphics[width=0.33\textwidth]{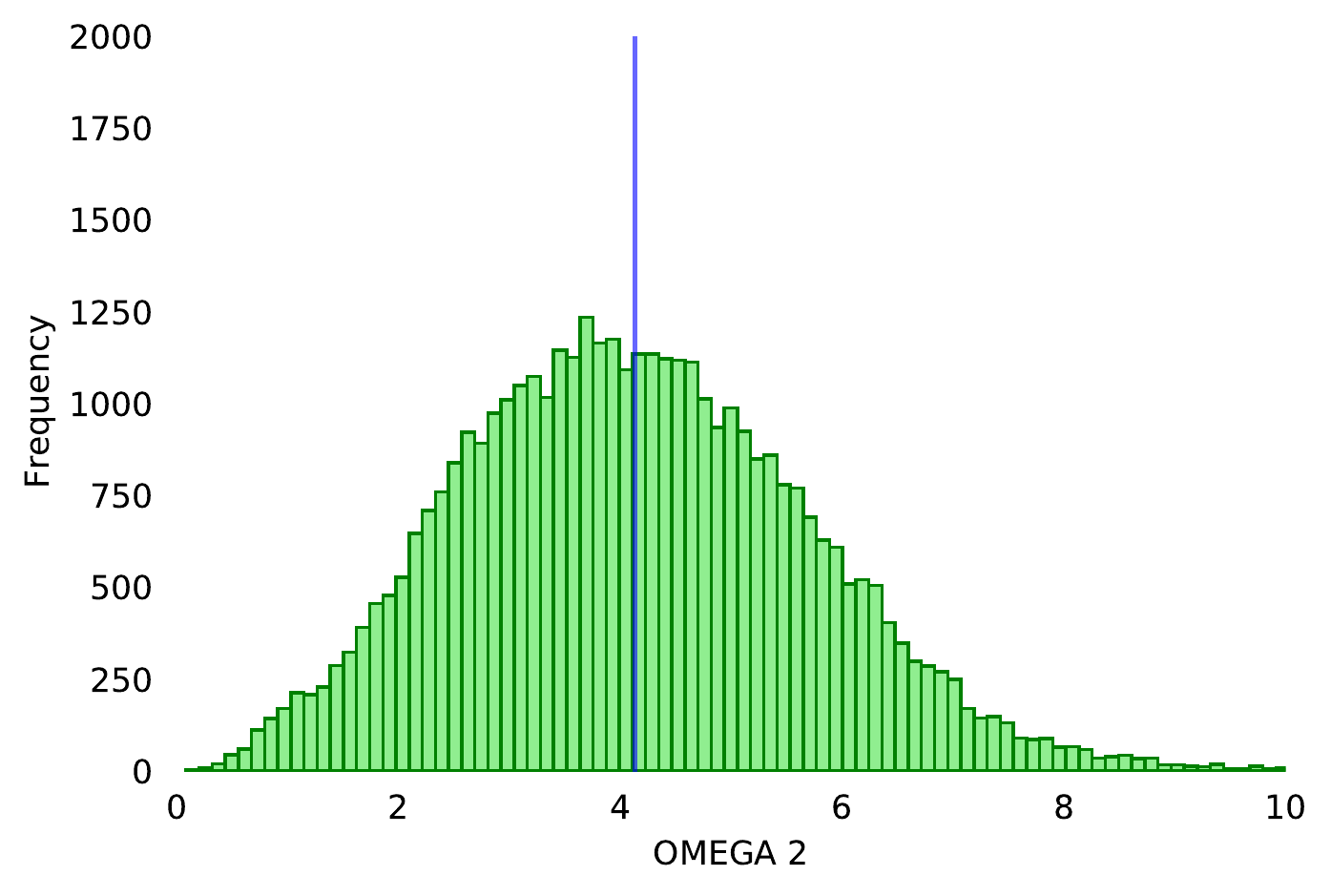}}
\hfill
\subfloat[\footnotesize $\Omega_3$  (25.323, 13.045)\label{fig:OMEGA3_moment_data_tumor_Matlab}]{\includegraphics[width=0.33\textwidth]{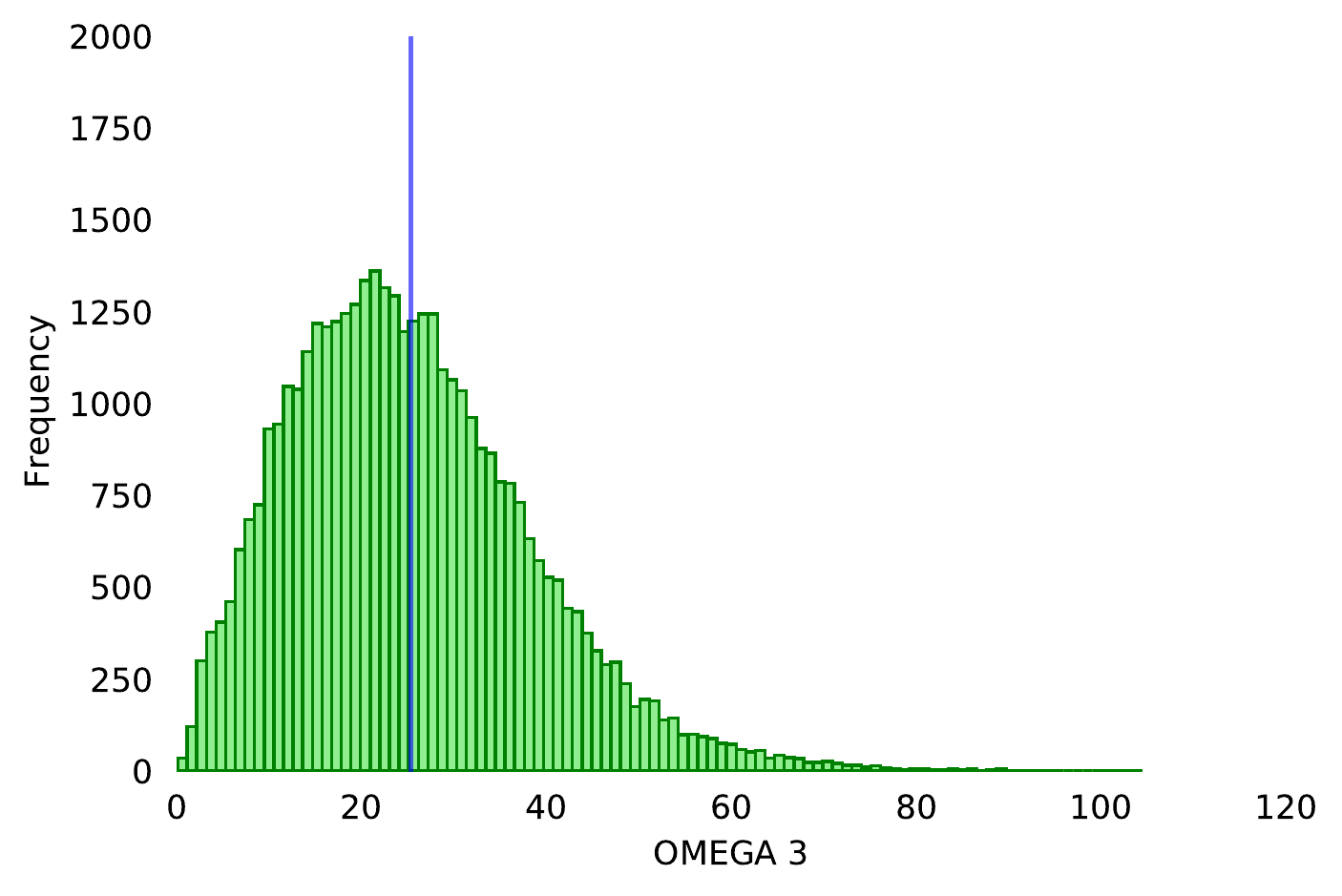}}
\caption{Moment invariants of generated data for the improved $\alpha$-GAN over 40,000 samples {(blue line represents the mean value)}}
\label{fig:moment_data_tumor_Matlab_sample_40000}
\end{figure}

The distribution of the 3D moment invariants $(\Omega_1, \Omega_2, \Omega_3)$ for generated samples are illustrated in Figure~\ref{fig:moment_data_tumor_Matlab_sample_40000}. The mean and standard deviation values show that the generated data consistently exhibit a higher mean and standard deviation. It seems like the diverse shapes of connected 3D tumors and lower levels of convexity compared to the spheres help the model generate more realistic-looking samples.

\begin{figure}[!ht]
    \centering
    \includegraphics[width=0.8\textwidth]{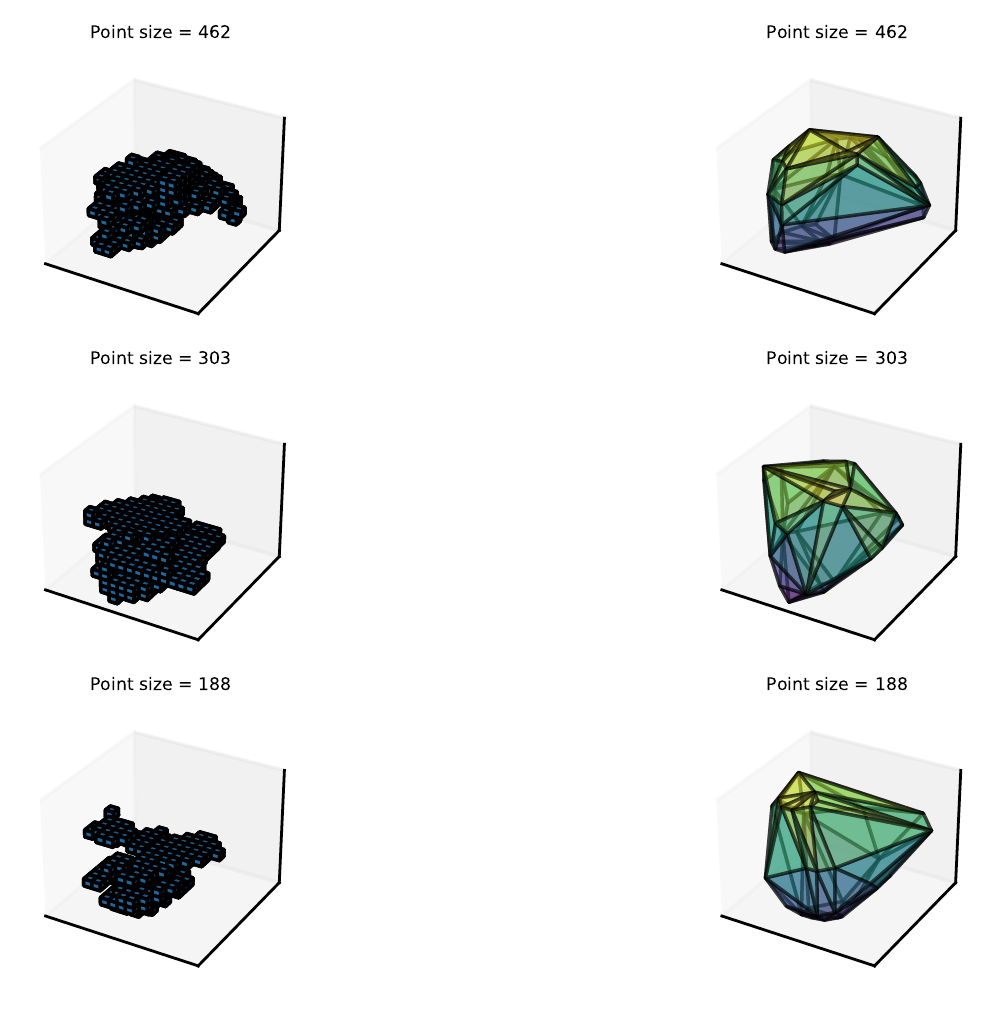}
    \caption{Voxel and mesh representation of sample generated tumor volumes' shapes for the improved $\alpha$-GAN {(sizes of the shapes are shown in the title of images)}}
    \label{fig:Matlab_gen_sample_trisurf}
\end{figure}

A sample of generated shapes illustrated in Figure~\ref{fig:Matlab_gen_sample_trisurf} shows the difference between the two datasets. The samples provide a clear look into the irregularities of a real volume's shape.

\subsubsection{Results with synthetic 3D connected volume with filled subspheres}

We report the shape and distance metrics for 3D connected volumes with packed spheres in Table~\ref{tbl:syntetic_results_tumor_iso_shape}. The improved $\alpha$-GAN structure shows a clear advantage in generating more connected subspheres. The low value of 1.170 for volume size KL divergence shows the capability of improved $\alpha$-GAN in generating connected shapes. This result attests to the ability of the model in generating representative 3D invariants in connected volumes as well. The connected loss has shown a great impact in generating more connected subspheres per shape. However, we also observe that the Shannon index depicts that subspheres' sizes are not uniform as training data, particularly for improved $\alpha$-GAN. 

\setlength{\tabcolsep}{5pt}
\renewcommand{\arraystretch}{1.6}
\begin{table}[!ht]
\caption{{\footnotesize Summary of performance values for the connected 3D volumes plus subspheres averaged over 5 repeats }}
\label{tbl:syntetic_results_tumor_iso_shape}
\begin{center}
\subfloat[Shape metrics]{
\resizebox{1.0\linewidth}{!}{
\begin{tabular}{lccllll}
\toprule
& & & \multicolumn{4}{c}{KL divergence} \\     
\cmidrule(lr){4-7} 
\mcps{\rrt GAN\\ model} & \mcps{\cnt Connected\\ volumes' sizes} & \mcps{\cnt Connected \\subspheres\\ per object} & \mcps{\cnt Connected\\ volumes' sizes} & \mcps{\cnt Shannon\\ index} & \mcps{\cnt Connectivity\\ ratio} & \mcps{\cnt Convexity\\ ratio}\\

  \midrule
$\alpha$-GAN &  278 $\pm$ 76 & 0.001 $\pm$ 0.016   & 11.289 $\pm$ 0.682 & 10.916 $\pm$ 0.682&  0.363  $\pm$ 0.088 & 11.499  $\pm$ 0.690 \\ 
$\alpha$-GAN-CL$^\dagger$ & 352 $\pm$ 109 &  0.013 $\pm$ 0.042 & 9.826  $\pm$ 0.662 &6.990  $\pm$ 0.667& 0.386  $\pm$0.091 & 11.358  $\pm$ 0.673\\ 
$\alpha$-GAN++$^\ddagger$ & 1,240 $\pm$ 257 & 0.073  $\pm$ 0.110 & 1.170   $\pm$ 0.171 & 11.057  $\pm$   0.687& 0.353  $\pm$ 0.076 & 11.246  $\pm$  0.670\\ 
\bottomrule 
\end{tabular}
}
}\\
\subfloat[Distance metrics]{
\resizebox{0.72\linewidth}{!}{
\begin{tabular}{lllll}
\toprule
&&  \multicolumn{3}{c}{KL divergence} \\     
\cmidrule(lr){3-5} 
\mcps{\rrt GAN\\ model} & \mcps{\cnt Subspheres' \\ coverage} & \mcps{\cnt FD error} & \mcps{\cnt Ratio \\ MAE} & \mcps{\cnt Target \\ distance error} \\
  \midrule
$\alpha$-GAN & 0.724  $\pm$ 0.058 & 4.176  $\pm$  0.499& 0.072  $\pm$  0.056& 14.187  $\pm$  2.075 \\   
$\alpha$-GAN-CL$^\dagger$ & 0.675 $\pm$ 0.058 & 4.371 $\pm$ 0.475 & 0.083  $\pm$ 0.062 & 14.844 $\pm$ 1.785\\ 
$\alpha$-GAN++$^\ddagger$ & 0.389  $\pm$  0.066 & 4.674  $\pm$ 0.528 &  0.122  $\pm$ 0.055 & 12.462  $\pm$  1.986\\ 
\bottomrule   
\end{tabular}
}
}
\\
{\tiny $^\dagger$: $\alpha$-GAN with connected loss,  $^\ddagger$: Improved $\alpha$-GAN}
\end{center}
\end{table}
Analyzing the distance metrics shows that subspheres' coverage decreases for improved $\alpha$-GAN. We find that the connected subspheres are more important than the subspheres' coverage. A set of unconnected points for one subsphere can cover a higher percentage of the connected volume. However, the generated shape does not qualify as suitable realistic-looking synthetic data. Therefore, we note that shape and distance metrics should jointly be considered for a model's evaluation. The KL divergence for distance metrics shows that improved GAN outperforms the other two variants in target distance error.

\begin{figure}[!ht]
\centering
\subfloat[Sample 1 with 7 subspheres \label{fig:sample1_tumor_iso_sphere}]{\includegraphics[width=1\textwidth]{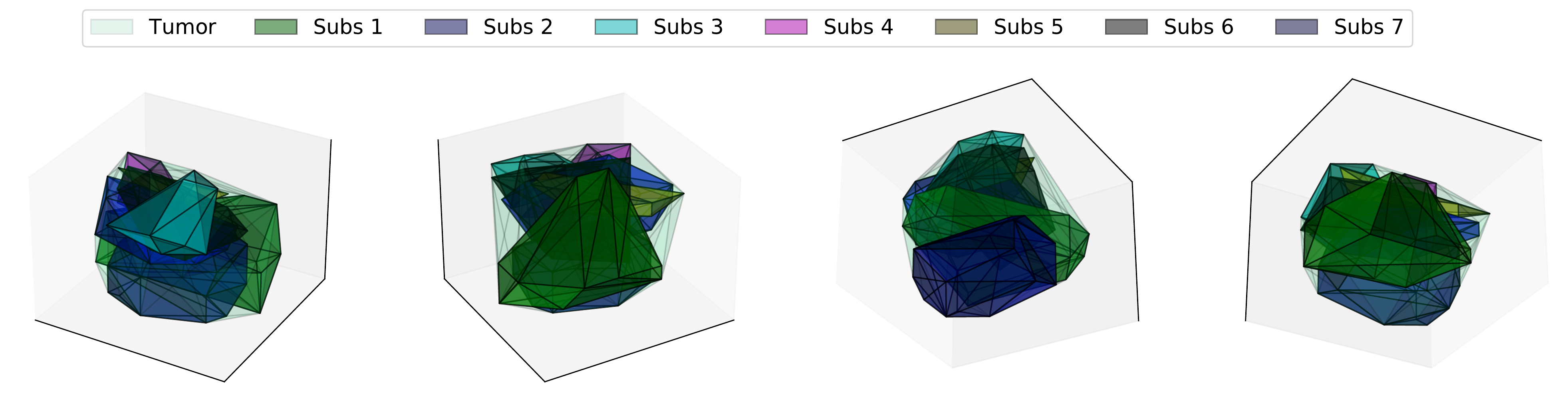}}
\hfill
\subfloat[Isocenters as the centers of the subspheres \label{fig:sample1_tumor_iso_sphere_subcenter}]{\includegraphics[width=1\textwidth]{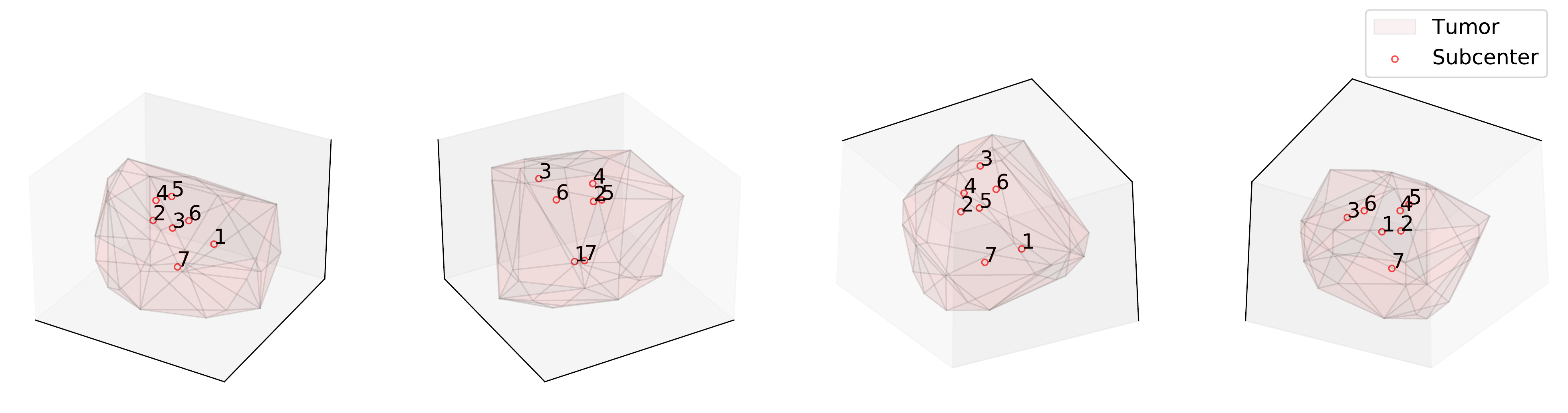}}
\hfill
\subfloat[Sample 2 with 7 subspheres \label{fig:sample2_tumor_iso_sphere}]{\includegraphics[width=1\textwidth]{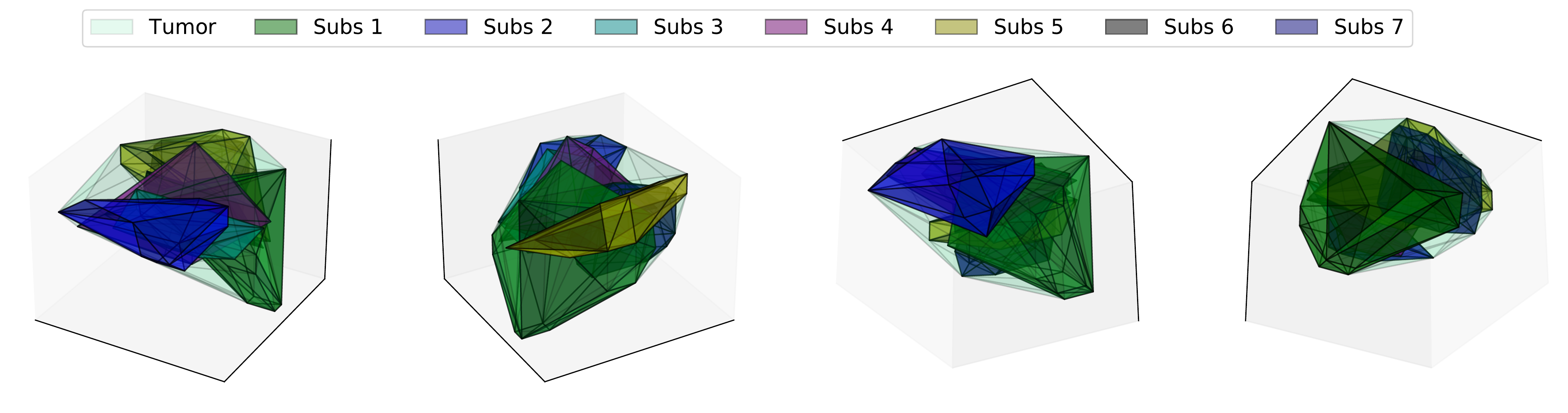}}
\hfill
\subfloat[Isocenters as the centers of the subspheres \label{fig:sample1_tumor_iso_sphere_subcenter}]{\includegraphics[width=1\textwidth]{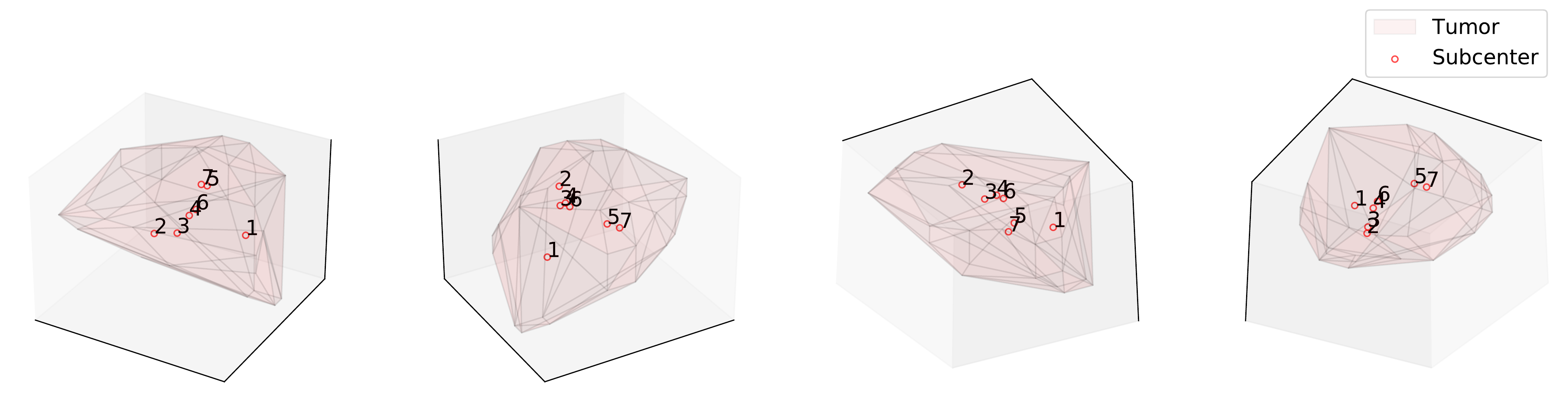}}

\caption{Sample of generated volumes plus subspheres results and relative isocenters for the improved $\alpha$-GAN {(different angles are illustrated)}}
\label{fig:sample_tumor_iso_results}
\end{figure}

A sample generated 3D connected volume shape and its subspheres are illustrated in Figure~\ref{fig:sample_tumor_iso_results}. Samples represent two volumes with different sizes and their respective subspheres. While the improved $\alpha$-GAN structure provides a higher percentage of connected subspheres per object compared to other models, the percentage is still smaller than expected. Therefore, a large portion of presented isocenters does not fit to the correct predefined coordinates. 


We report the shape evaluation metrics for each generated  subsphere in Table~\ref{tbl:results_tumor_iso_matlab_shape}. The improved $\alpha$-GAN architecture shows high values of connected subspheres per connected volume, confirming the model's ability in generating highly connected 3D objects. As expected, the subspheres' coverage ratio is lower than the other two variants. On the other hand, in contrast to the result for other datasets, improved $\alpha$-GAN shows great resistance to the variability of the number of isocenters and their subspheres' irregular shapes and outperforms others in terms of the KL divergence of evaluation metrics. Note that due to the variant sizes of subspheres, KL divergence is reported rather than the Shannon index.

\setlength{\tabcolsep}{6pt}
\renewcommand{\arraystretch}{2}
\begin{table}[!ht]
\caption{{\footnotesize Summary of performance values for the connected 3D tumors with subspheres averaged over 5 repeats }}
\label{tbl:results_tumor_iso_matlab_shape}
\begin{center}
\resizebox{1.0\linewidth}{!}{
\begin{tabular}{lccccllll}
\toprule
& & & &&\multicolumn{4}{c}{KL divergence} \\     
\cmidrule(lr){6-9} 
\mcps{\rrt GAN\\ model} & \mcps{\cnt Connected\\ volumes' sizes} & \mcps{\cnt Connected \\subspheres\\ per object} & \mcps{\cnt Subspheres' \\ coverage} &&\mcps{\cnt Connected\\ volumes' sizes} & \mcps{\cnt Shannon\\ index} & \mcps{\cnt Connectivity\\ ratio} & \mcps{\cnt Convexity\\ ratio}\\
\midrule

$\alpha$-GAN & 149 $\pm$ 49 & 0.001 $\pm$ 0.007 & 0.616 $\pm$ 0.095 &  & 2.369 $\pm$ 0.332 &  0.910 $\pm$ 0.233& 0.456 $\pm$ 0.138 & 10.111 $\pm$  0.551\\
$\alpha$-GAN-CL$^\dagger$ & 148 $\pm$ 50&   0.006 $\pm$ 0.025 & 0.641 $\pm$ 0.091 & &2.537 $\pm$ 0.336 & 0.476 $\pm$  0.120 & 0.428 $\pm$ 0.128 & 9.863 $\pm$ 0.532 \\ 
$\alpha$-GAN++$^\ddagger$ & 283 $\pm$ 70 & 0.130 $\pm$ 0.132 & 0.531 $\pm$ 0.153 && 0.446 $\pm$ 0.067 & 0.485 $\pm$ 0.120 & 0.402 $\pm$ 0.118 &  6.570 $\pm$ 0.320\\

\bottomrule         
\end{tabular}
}
\\
{\tiny $^\dagger$: $\alpha$-GAN with connected loss,  $^\ddagger$: Improved $\alpha$-GAN}
\end{center}
\end{table}

\begin{figure}[!ht]
\centering
\subfloat[Sample with 3 subspheres \label{fig:sample1_tumor_iso_shape}]{\includegraphics[width=1\textwidth]{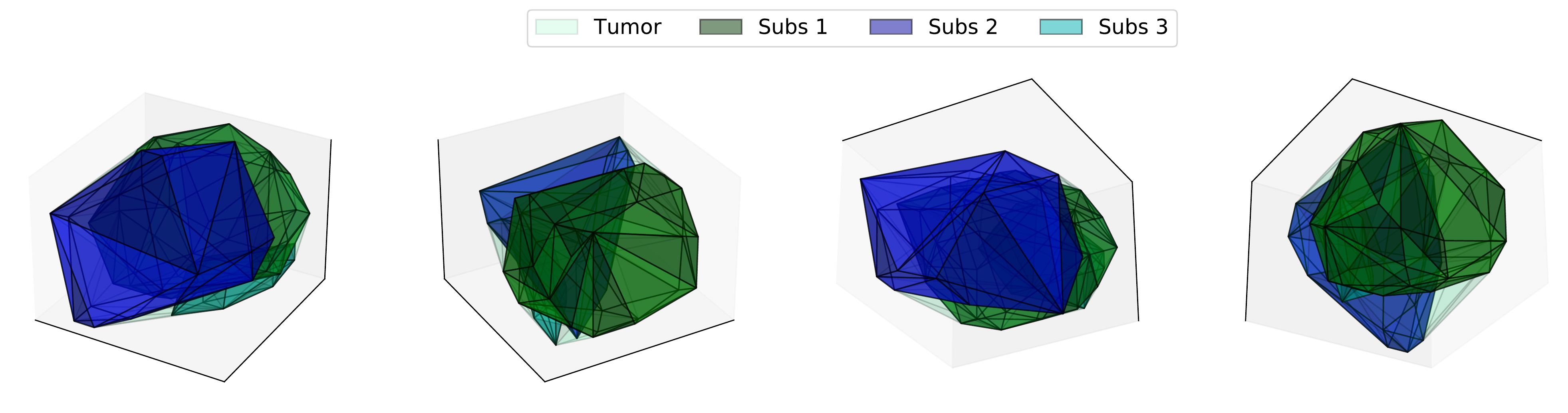}}
\hfill
\subfloat[Isocenters as the centers of the subspheres \label{fig:sample1_tumor_iso_subcenters}]{\includegraphics[width=1\textwidth]{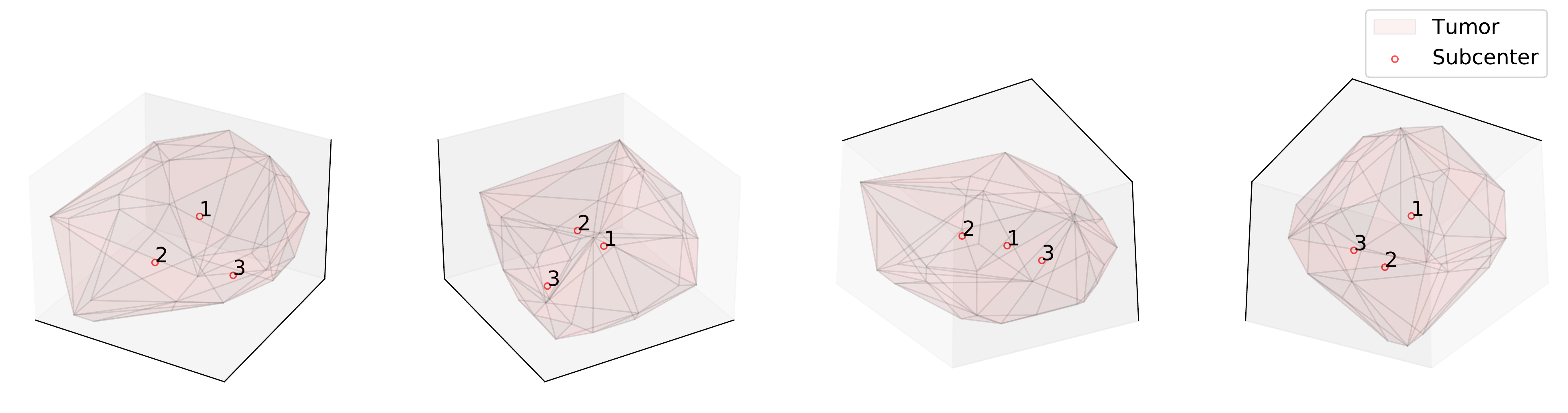}}
\hfill
\subfloat[Sample with 3 subspheres \label{fig:sample2_tumor_iso_shape}]{\includegraphics[width=1\textwidth]{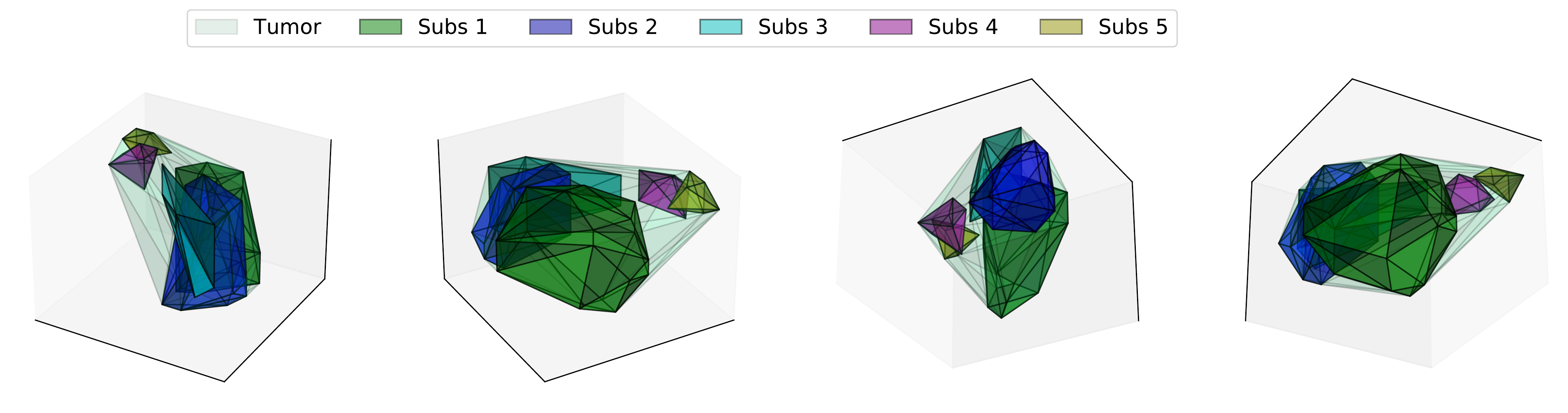}}
\hfill
\subfloat[Isocenters as the centers of the subspheres \label{fig:sample2_tumor_iso_subcenters}]{\includegraphics[width=1\textwidth]{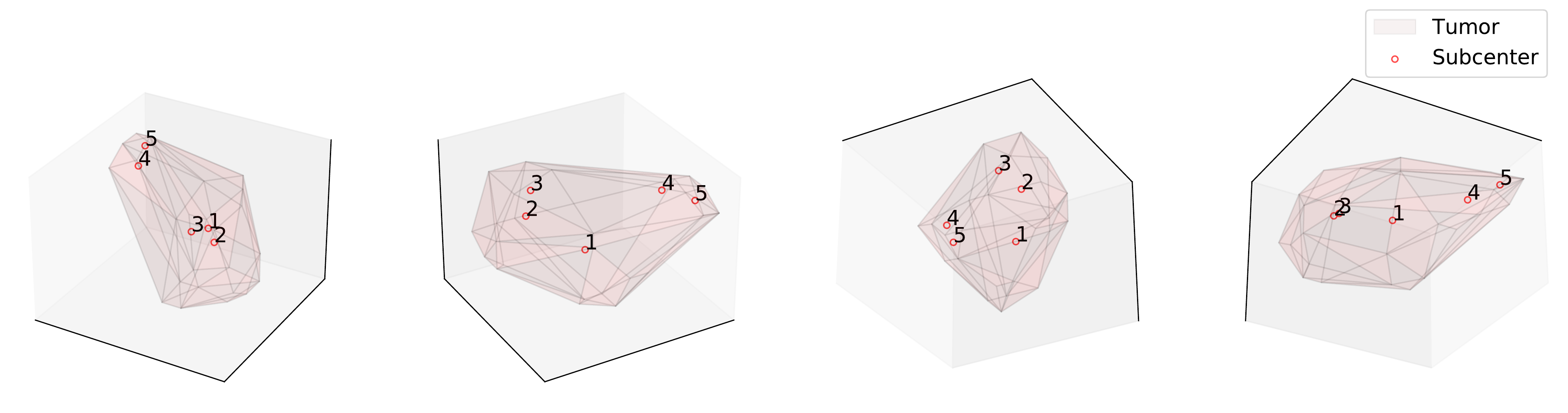}}
\caption{Sample generated volumes with subspheres and relative isocenters for the improved $\alpha$-GAN {(different angles are illustrated)}}
\label{fig:Matlab_gen_iso_sample_defined}
\end{figure}

A sample 3D connected tumor volume along with its relevant subspheres are illustrated in Figure~\ref{fig:Matlab_gen_iso_sample_defined}. We observe that sample connected volumes present different numbers of subspheres with diverse sets of radiuses. The location of generated isocenters shows a reasonable distribution over the tumor space.



\section{Discussions and conclusions}\label{sec:conclusion}
Our numerical study demonstrates the capabilities and limitations of 3D GAN architectures in generating 3D connected volumes. 
In particular, we note that the addition of predefined subspheres to the connected volumes further complicates the problem.
As such, it is deemed essential to investigate custom 3D GAN architectures for this task.
We observe that $\alpha$-GAN outperforms the baseline GAN architectures in generating 3D connected volumes. However, by involving the subspheres and increasing the data diversity, $\alpha$-GAN fails to maintain the connectivity within 3D shape and subspheres. 
Our numerical study shows that improved $\alpha$-GAN is more resilient to variable subsphere sizes, and outperforms the other models in volume plus subsphere generation for tumor synthetic data. Our proposed modifications in the discriminator's network lead to a comparable performance with the $\alpha$-GAN in generating connected shapes. 
More importantly, it shows more realistic-looking samples compared to $\alpha$-GAN. 

We note that preserving the connectivity of the generated shapes is one of the main goals of this study. 
Therefore, the presented results reveal an important finding in this regard, as they show the impact of inception layers inside the discriminator in detecting the connectivity between the 3D voxels.
The methods and results obtained in this work can benefit further research on radiation therapy treatment planning by generating high-quality synthetic datasets since the connected volumes share many similarities with the tumors. 
That is, the connectivity between tumor voxels is a crucial aspect that can be preserved using our proposed GAN architecture. 

Our proposed 3D $\alpha$-GAN  architecture can also handle connected 3D volumes together with their subspheres. 
We examine two baseline GAN models along with $\alpha$-GAN variants on four different synthetically generated datasets. Experimental results reveal that $\alpha$-GAN models outperform the baseline GAN models in generating connected 3D volumes. 
More importantly, the results of our proposed architecture on volumes with subspheres highlight the value of having small kernels in the inception layers to better detect the connectivity features.
In this regard, the ability of our approach to generating 3D tumor volumes with designated isocenter locations (i.e., as subspheres) has the potential to significantly contribute to research on radiosurgery treatment planning as it opens the door for open sourcing realistic-looking datasets.

There are several future research directions for this work.
In this study, we limited our approach to generating one connected volume in each 3D object. 
In future research, we plan to investigate the effect of additional connected volumes on the connectivity of samples. 
In addition, we only focused on tumor volumes as connected 3D shapes and investigating the applications of 3D connected volumes in other domains is left for future research.

\section*{Data availability statement}
All the datasets are publicly available, and can be obtained using the described methods.

\section*{Disclosure statement}
No potential conflict of interest was reported by the authors.

 \bibliographystyle{spbasic} 
 \bibliography{refs}

\newpage
\appendix

\section{Model architectures}\label{ap:model_arch}

Table~\ref{tbl:alpha_GAN_model_struct} shows the $\alpha$-GAN architecture adopted from \citep{kwon2019generation}. 

\setlength{\tabcolsep}{3pt}
\renewcommand{\arraystretch}{1.1}
\begin{table}[!ht]
\caption{{Structure of  $\alpha$-GAN model}}
\label{tbl:alpha_GAN_model_struct}
\begin{center}
\resizebox{1\linewidth}{!}{
\noindent\adjustbox{max width=\textwidth}{

\begin{tabular}{lccccc}
\toprule
 Network &Layer &  \# of channels & Kernel size& Activation function & Batch normalization\\
\toprule

\multirow{4}{*}{Generator}
& Conv & 16 & 4 & ReLU & True\\
& Upsample & - & - & - & False\\
&  Conv & 8  & 4& ReLU & True\\
& Upsample & - & - & - & False\\
& Conv & 1 & 4 & Sigmoid & False\\
\toprule
\multirow{5}{*}{Discriminator} 
&  Conv & 4 & 4 & LeakyReLU & False\\
& Conv & 8 & 4 & LeakyReLU & True\\
& Conv& 16 & 4 & LeakyReLU & True\\
& Conv & 1 & 4 & LeakyReLU & False\\
\toprule
\multirow{5}{*}{Code Discriminator} & & units & & \\
\midrule
&  Linear & 4096 & - & LeakyReLU & True\\
& Linear & 4096 & - & LeakyReLU & True\\
& Linear& 1 & - & - & False \\

\toprule
\end{tabular}}

}
\end{center}
\end{table}

Table~\ref{tbl:model_struct_improved} presents the details of  improved $\alpha$-GAN discriminator's architecture using the inception block showed in Table~\ref{tbl:inception_model_struct_improved}. The other networks have a similar structure as Table~\ref{tbl:alpha_GAN_model_struct}.

\setlength{\tabcolsep}{3pt}
\renewcommand{\arraystretch}{1.1}
\begin{table}[!ht]
\caption{{Structure of inception block used in improved $\alpha$-GAN structure}}
\label{tbl:inception_model_struct_improved}
\begin{center}
\resizebox{1\linewidth}{!}{
\noindent\adjustbox{max width=\textwidth}{

\begin{tabular}{cccccc}
\toprule
 Network &Layer &  \# of channels & Kernel size& Activation function & Batch normalization\\
\toprule
\multirow{2}{*}{Block 1} 
&  Conv & $n_{in}/2$ & 1 & ReLU & True\\
& Conv & $n_{out}$& 5 & ReLU & True\\
\midrule
\multirow{2}{*}{Block 2} & Conv& $n_{in}/2$ & 1 & ReLU & True\\
& Conv & $n_{out}$& 3 & ReLU& True\\
\midrule
\multirow{1}{*}{Block 3} & Conv& $n_{out}$ & 1 & ReLU & True\\
\midrule
\multirow{2}{*}{Block 4} & MaxPool& - & - & ReLU & False\\
& Conv & $n_{out}$& 1 & ReLU& True\\
\midrule
\multirow{1}{*}{Merge} & $n_{out}$ $\times$ 4 & - & - & -\\
\toprule
\end{tabular}}

}
\end{center}
\end{table}

\setlength{\tabcolsep}{3pt}
\renewcommand{\arraystretch}{1.4}
\begin{table}[!ht]
\caption{{ Structure of improved $\alpha$-GAN model}}
\label{tbl:model_struct_improved}
\begin{center}
\resizebox{1\linewidth}{!}{
\noindent\adjustbox{max width=\textwidth}{

\begin{tabular}{cccccc}
\toprule
 Network &Layer &  \# of channels & Kernel size& Activation function & Batch normalization\\
\toprule
\multirow{3}{*}{Discriminator} 
&  Inception & 4 & - & - & -\\
& Inception & 16 & - & - & -\\
& Linear & - & - &  &-\\
\toprule
\end{tabular}}

}
\end{center}
\end{table}


\newpage
\section{Notations}\label{ap:notations}
We provide a summary of mathematical notations used in the paper in Table~\ref{tbl:notations}.
\setlength{\tabcolsep}{9pt}
\renewcommand{\arraystretch}{1.3}

\begin{table}[!ht]
    \centering
    \small
    \caption{Mathematical notations}
    \label{tbl:notations}
    \resizebox{1\linewidth}{!}{
    \begin{tabularx}{\textwidth}{lX}
    \toprule
    Notation &  Description\\
    \midrule
    $G(z;\theta_g)$ & Generator network of GAN with $z$ input and $\theta_g$ set of parameters\\
    $D(x;\theta_d)$ & Discriminator network of GAN with $x$ input and $\theta_d$ set of parameters\\
     $D_c(x;\theta_c)$ & Code discriminator of $\alpha$-GAN with $x$ input and $\theta_c$ set of parameters\\
      $E(x;\theta_e)$ & Encoder network of $\alpha$-GAN with $x$ input and $\theta_e$ set of parameters\\
         $\hat{x}$ & Reconstructed version of $x$ generated by an auto-encoder\\
         $\mathcal{L}(x,\hat{x})$ & Reconstruction loss between the $x$ and $\hat{x}$\\
         $z_e$ & Decoded noise vector from real samples\\
         $z_r$ & Real noise input vector\\
         $L_{\texttt{GPD}}$ & Gradient penalty loss for the discriminator\\
                  $L_{\texttt{GPC}}$ & Gradient penalty loss for the code discriminator\\
        
        $\texttt{CC}$ & Connected components \\
                  
    \bottomrule
    \end{tabularx}}

\end{table}

\end{document}